\documentclass{jpsj3}
\usepackage{type1cm}
\usepackage{tabularx}
\usepackage[dvipdfmx]{color}

\title{
Theoretical Study on 
Anisotropic Magnetoresistance Effects 
of ${\mbox{\boldmath $I$}}//[100]$, 
${\mbox{\boldmath $I$}}//[110]$, 
and ${\mbox{\boldmath $I$}}//[001]$ 
for Ferromagnets 
with a Crystal Field of Tetragonal Symmetry
}
\author{Satoshi Kokado$^1$\thanks{E-mail address: 
kokado.satoshi@shizuoka.ac.jp
} and Masakiyo Tsunoda$^{2,3}$ 
}
\inst{
$^1$Electronics and Materials Science Course, 
Department of Engineering, 
Graduate School of Integrated Science and Technology, 
Shizuoka University, Hamamatsu 432-8561, Japan \\
$^2$Department of Electronic Engineering, 
Graduate School of Engineering, Tohoku University, 
Sendai 980-8579, Japan\\
$^3$Center for Spintronics Research Network, Tohoku University, 
Sendai 980-8577, Japan
}
\abst{
Using the electron scattering theory, 
we obtain analytic expressions for 
anisotropic magnetoresistance (AMR) ratios 
for ferromagnets with a crystal field of tetragonal symmetry. 
Here, 
a tetragonal distortion exists in the [001] direction, 
the magnetization ${\mbox{\boldmath $M$}}$ 
lies in the (001) plane, 
and the current ${\mbox{\boldmath $I$}}$ flows 
in the [100], [010], or [001] direction. 
When 
the ${\mbox{\boldmath $I$}}$ direction is denoted by $i$, 
we obtain the AMR ratio 
as 
${\rm AMR}^i (\phi_i)=
C_0^i + C_2^i \cos 2\phi_i + C_4^i \cos 4 \phi_i \ldots = \sum_{j=0,2,4,\ldots} C_j^i \cos j\phi_i$, 
with $i=[100]$, $[110]$, and $[001]$, 
$\phi_{[100]} = \phi_{[001]}=\phi$, and $\phi_{[110]}=\phi'$. 
The quantity 
$\phi$ ($\phi'$) is the relative angle between 
${\mbox{\boldmath $M$}}$ 
and the $[100]$ ($[110]$) direction, 
and $C_j^i$ is a coefficient composed of 
a spin--orbit coupling constant, an exchange field, 
the crystal field, and resistivities. 
We elucidate 
the origin of $C_j^i \cos j\phi_i$ 
and 
the features of $C_j^i$. 
In addition, we obtain 
the relation $C_4^{[100]} = -C_4^{[110]}$, 
which was experimentally observed for Ni, 
under a certain condition. 
We also qualitatively explain 
the experimental results of 
$C_2^{[100]}$, $C_4^{[100]}$, $C_2^{[110]}$, and $C_4^{[110]}$ at 293 K 
for Ni.

}

\begin{document}
\maketitle

\section{Introduction}
The anisotropic magnetoresistance (AMR) effect for ferromagnets,\cite{Thomson,McGuire,Campbell,Potter,Miyazaki,Tsunoda1,Tsunoda2,Tsunoda3,Kabara1,Ito,ZRLi,Takata,Oogane,Gorkom,Yang,Sakuraba,Heusler,Liu,Du,Ueda,Nishiwaki,Yako,
Miyakozawa,Zhao,Kokado1,Kokado2,Kokado3,Kokado3_1,Kokado4,Dedie} 
in which the electrical resistivity depends on 
the direction of magnetization ${\mbox{\boldmath $M$}}$, 
has been studied extensively both experimentally and theoretically. 
The efficiency of the effect ``AMR ratio'' 
is defined by
\begin{eqnarray}
\label{AMR^i(phi_i)}
{\rm AMR}^i(\phi_i)
= \frac{\rho^i (\phi_i) - \rho_\perp^i}{\rho_\perp^i}, 
\end{eqnarray}
with 
$\rho_\perp^i=\rho^i (\pi/2)$. 
Here, 
$\rho^i (\phi_i)$ is the resistivity at $\phi_i$ 
in the current ${\mbox{\boldmath $I$}}$ direction, $i$, 
where 
$\phi_i$ is the relative angle between 
the thermal average of 
the spin $\langle {\mbox{\boldmath $S$}} \rangle$ 
($\propto$$-{\mbox{\boldmath $M$}}$) 
and a specific direction 
for the case of $i$.

The AMR ratio 
AMR$^i$ (0) has often been investigated for many magnetic materials. 
In particular, the experimental results of AMR$^i$(0) 
for Ni-based alloys 
have been analyzed 
by using the electron scattering theory 
with no crystal field, i.e., 
the Campbell--Fert--Jaoul (CFJ) model\cite{Campbell}. 
We have recently extended this CFJ model to a general model 
that can qualitatively explain AMR$^i$(0) 
for various ferromagnets\cite{Kokado1,Kokado2}.

On the other hand, 
when 
$\langle {\mbox{\boldmath $S$}} \rangle$ 
lies in the (001) plane 
and ${\mbox{\boldmath $I$}}$ flows 
in the $i$ direction, 
with $i=[100]$ and $[110]$, 
AMR$^i$($\phi_i$) has been experimentally observed 
to be\cite{Tsunoda2,Tsunoda3,Kabara1,Ito,ZRLi,Takata,Oogane,Gorkom} 
\begin{eqnarray}
\label{AMR(phi)}
&&{\rm AMR}^i(\phi_i)
= C_0^i + C_2^i \cos 2\phi_i + C_4^i \cos 4\phi_i
+ \ldots \\
\label{AMR(phi)1}
&&\hspace{1.9cm}=\sum_{j=0,2,4,\ldots} C_j^i \cos j \phi_i, 
\end{eqnarray}
with $\phi_{[100]} = \phi$ 
and $\phi_{[110]}=\phi'$, 
where $\phi$ is the relative angle 
between the $\langle {\mbox{\boldmath $S$}} \rangle$ direction 
and the [100] direction (see Fig. \ref{sample}) 
and $\phi'$ is the relative angle 
between the $\langle {\mbox{\boldmath $S$}} \rangle$ direction 
and the [110] direction (see Fig. \ref{sample}). 
In addition, 
$C_0^i$ is 
the constant term 
in the case of $i$, 
and 
$C_j^i$ is the coefficient of 
the $\cos j\phi_i$ term 
in the case of $i$. 
The case of Eq. (\ref{AMR(phi)}) with $C_2^i \ne 0$ and $C_j^i=0$ ($j \ge 4$) 
is called the twofold symmetric AMR effect, 
while 
the case of Eq. (\ref{AMR(phi)}) with $C_2^i \ne 0$ 
and $C_j^i \ne 0$ 
($j \ge 4$) 
is 
the higher-order fold symmetric AMR effect. 
The twofold symmetric AMR effect 
has often been observed for various ferromagnets 
and 
analyzed 
on the basis of our previous model\cite{Kokado1,Kokado2}. 
The higher-order fold symmetric AMR effect 
of ${\mbox{\boldmath $I$}}//[100]$ and ${\mbox{\boldmath $I$}}//[110]$
has been observed for typical ferromagnets 
Ni\cite{Dedie,Ni_C4}, 
Fe$_4$N\cite{Tsunoda2}, and Ni$_x$Fe$_{4-x}$N ($x=1$ and 3)\cite{Takata}. 
In particular, 
the relation 
\begin{eqnarray}
\label{relation}
C_4^{[100]}=-C_4^{[110]} 
\end{eqnarray}
has been found 
in the temperature dependence 
of the AMR ratio.\cite{Dedie,Ni_C4,Tsunoda2,Takata}

The AMR ratio of Eq. (\ref{AMR(phi)}) 
has sometimes been fitted by 
using an expression by D$\ddot{\rm o}$ring. 
This expression consists of an expression for the resistivity, 
which is based on the symmetry of a crystal 
(see Appendix \ref{appen_pheno}).\cite{Doring,Bozorth,Gorkom} 
D$\ddot{\rm o}$ring's expression can be easily applied to 
the cases of 
the arbitrary directions of 
${\mbox{\boldmath $I$}}$ and ${\mbox{\boldmath $M$}}$. 
The expression, however, 
has been considered unsuitable for 
physical consideration 
because it was not based on the electron scattering theory.

To improve this situation, 
we have recently developed a theory of 
the twofold and fourfold symmetric AMR effect 
using the electron scattering theory. 
Here, we derived 
an expression for 
${\rm AMR}^{[100]}(\phi)$ of Eq. (\ref{AMR(phi)}) 
for ferromagnets with a crystal field. 
As a result, we found that 
$C_4^{[100]}$ appears under a crystal field of tetragonal symmetry, 
whereas 
it takes a value of almost 0 
under a crystal field of cubic symmetry\cite{Kokado3}. 
The expression for AMR$^{[110]}(\phi')$, 
however, 
has scarcely been derived.

In the future, 
not only the expression for ${\rm AMR}^{[100]}(\phi)$ but also 
expressions for ${\rm AMR}^{[110]}(\phi')$ and so on 
will play an important role 
in theoretical analyses and 
physical considerations of experimental results.
In addition, 
Eq. (\ref{relation}) should be confirmed 
by using the electron scattering theory.

In this paper, 
using the electron scattering theory, 
we first obtained analytic expressions for 
${\rm AMR}^i (\phi_i)$ of Eq. (\ref{AMR(phi)1}) 
for ferromagnets with a crystal field of tetragonal symmetry, 
where $i=[100]$, $[110]$, and $[001]$; 
$\phi_{[100]} = \phi_{[001]}=\phi$; and $\phi_{[110]}=\phi'$ 
(see Fig. \ref{sample}). 
Second, we elucidated 
the origin of $C_j^i \cos j \phi_i$ 
and 
the features of $C_j^i$. 
In addition, we obtained the relation $C_4^{[100]} = -C_4^{[110]}$ 
of Eq. (\ref{relation}) under a certain condition.
Third, we qualitatively explained 
the experimental result of $C_j^i$ at 293 K for Ni 
using the expression for $C_j^i$. 
The AMR ratios 
AMR$^{[100]}(0)$ and AMR$^{[110]}(0)$ 
also corresponded to 
that of 
the CFJ model\cite{Campbell} 
under the condition of the CFJ model.

The present paper is organized as follows. 
In Sect. \ref{sec_theory}, 
we present the electron scattering theory, 
which takes into account the localized d states 
with a crystal field of tetragonal symmetry. 
We first obtain wave functions of the d states 
using the first- and second-order perturbation theory. 
Second, we show the expression for the resistivity, 
which is composed of the wave functions of the d states. 
In Sect. \ref{sec_calc}, 
we describe the expressions for 
AMR$^i(\phi_i)$ 
for ferromagnets including half-metallic ferromagnets.
In Sect. \ref{sec_cons}, 
we elucidate 
the origin of $C_j^i \cos j \phi_i$ 
and 
the features of $C_j^i$. 
In Sect. \ref{C_4^100=-C_4^110}, 
the relation $C_4^{[100]} = -C_4^{[110]}$ is obtained 
under a certain condition. 
In Sect. \ref{sec_Ni}, 
we qualitatively explain 
the experimental result of $C_j^i$ at 293 K for Ni. 
The conclusion is presented in Sect. \ref{sec_conc}. 
In Appendix \ref{appen_pheno}, we report 
the expression for the AMR ratio 
by D$\ddot{\rm o}$ring. 
In Appendix \ref{appen_wf}, we 
give 
an expression for 
a wave function 
obtained by applying the perturbation theory 
to a model with degenerate unperturbed systems. 
In Appendix \ref{appen_resis}, 
we describe 
the expressions for resistivities for the present model. 
In Appendix \ref{appen_coeff}, 
$C_j^i$ is expressed as a function of the resistivities. 
In Appendix \ref{appen_C_j^i}, we give the expression for $C_j^i$. 
In Appendix \ref{appen_origin}, 
we explain the origin of $C_j^i \cos j \phi_i$. 
In Appendix \ref{appen_CFJ}, 
we show that 
the present model corresponds to the CFJ model\cite{Campbell} 
under the condition of the CFJ model.

\section{Theory}
\label{sec_theory}
In this section, we describe the electron scattering theory to obtain 
$\rho^i(\phi_i)$ 
and 
AMR$^i(\phi_i)$ with $i=[100]$, $[110]$, and $[001]$ for the ferromagnets.

\subsection{Model}
\label{sec_model}
\begin{figure}[ht]
\begin{center}
\includegraphics[width=0.6\linewidth]{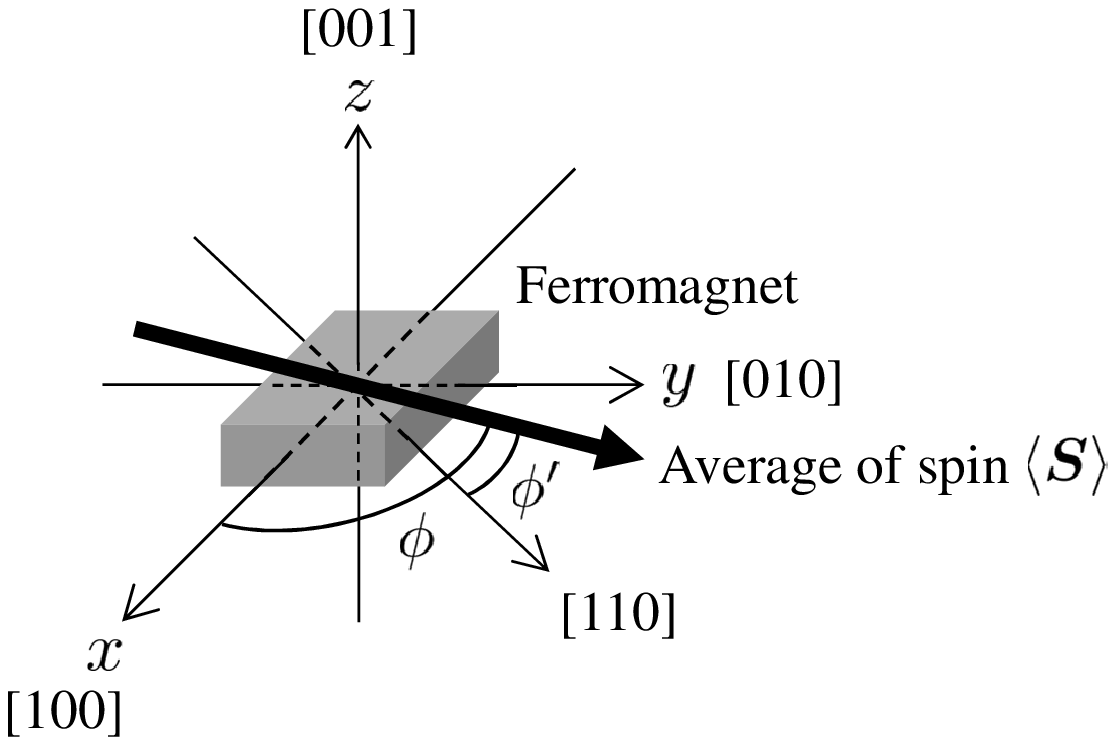}
\caption{
Sketch of the sample geometry. 
The tetragonal distortion is in the [001] direction. 
The current ${\mbox{\boldmath $I$}}$ flows 
in the [100], [110], or [001] direction. 
The thermal average of the spin 
$\langle {\mbox{\boldmath $S$}} \rangle$ 
($\propto$$-{\mbox{\boldmath $M$}}$) lies in the (001) plane. 
For $\phi_i$, 
we set $\phi_{[100]} = \phi_{[001]}=\phi$ and $\phi_{[110]}=\phi'$. 
Here, $\phi$ is the relative angle between 
$\langle {\mbox{\boldmath $S$}} \rangle$ and 
the [100] direction, 
and 
$\phi'$ is the relative angle between 
the $\langle {\mbox{\boldmath $S$}} \rangle$ direction 
and the [110] direction. 
The relation between $\phi$ and $\phi'$ 
is given by Eq. (\ref{phi_phi'}). 
Furthermore, the $x$-, $y$-, and $z$-axes are specified 
to describe the Hamiltonian of Eq. (\ref{Hamiltonian}). 
}
\label{sample}
\end{center}
\end{figure}

Figure \ref{sample} shows the present system, 
in which 
a tetragonal distortion exists in the [001] direction, 
the thermal average of the spin 
$\langle {\mbox{\boldmath $S$}} \rangle$ 
($\propto$$-{\mbox{\boldmath $M$}}$) 
lies in the (001) plane, 
and the current ${\mbox{\boldmath $I$}}$ flows 
in the [100], [010], or [001] direction. 
For $\phi_i$, 
we set $\phi_{[100]} = \phi_{[001]}=\phi$ and $\phi_{[110]}=\phi'$. 
The relation between $\phi$ and $\phi'$ 
is given by 
\begin{eqnarray}
\label{phi_phi'}
\phi=\phi' + \frac{\pi}{4}.
\end{eqnarray}

For this system, we use the two-current model 
with the $s$--$s$ and $s$--$d$ scatterings.\cite{Kokado1,Kokado2,Kokado3,Kokado3_1} 
The $s$--$s$ scattering represents 
the scattering of 
the conduction electron ($s$) 
into the conduction state ($s$) 
by nonmagnetic impurities and phonons. 
The $s$--$d$ scattering represents 
the scattering of 
the conduction electron ($s$) 
into the localized d states ($d$) by nonmagnetic impurities. 
Here, 
the conduction state 
consists of s, p, and the conductive d states. 
The localized d states are obtained 
by applying the perturbation theory 
to the Hamiltonian of the d states, ${\cal H}$.

\subsection{Hamiltonian}
\label{sub_Ham}
\begin{figure}[ht]
\begin{center}
\includegraphics[width=0.5\linewidth]{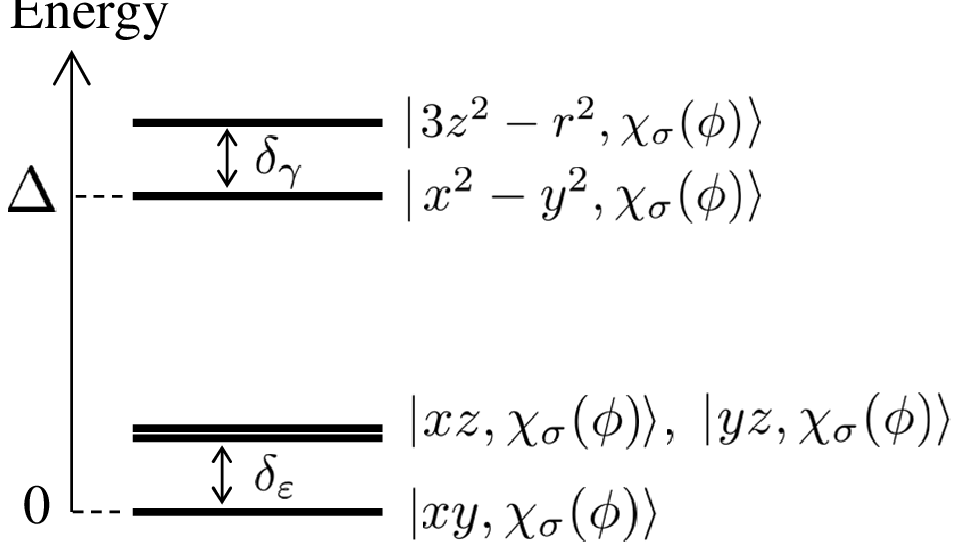}
\caption{
Energy levels of the 3d states 
in the crystal field of tetragonal symmetry.\cite{Yosida1} 
The second excited states are doubly degenerate. 
The energy levels are measured from 
the energy level of $|xy \rangle$, 
$E_\varepsilon$. 
}
\label{energy}
\end{center}
\end{figure}

Following our previous study\cite{Kokado3,Kokado3_1}, 
we consider ${\cal H}$ as the Hamiltonian of the localized d states 
of a single atom in a ferromagnet with 
a spin--orbit interaction, an exchange field, 
and a crystal field of tetragonal symmetry. 
This crystal field represents 
the case that a distortion in the [001] direction is added to 
a crystal field of cubic symmetry. 
The reason for choosing this crystal field 
is that 
$C_4^{[100]}$ appears under the crystal field of tetragonal symmetry, 
whereas 
it takes a value of almost 0 
under the crystal field of cubic symmetry, 
as reported in Refs. \citen{Kokado3} and \citen{others}. 
The Hamiltonian ${\cal H}$ is expressed as
\begin{eqnarray}
\label{Hamiltonian}
&&{\cal H} = {\cal H}_0 + V, \\
\label{Hamiltonian0}
&&{\cal H}_0 = {\cal H}_{\rm cubic} 
- {\mbox{\boldmath $S$}} \cdot {\mbox{\boldmath $H$}}, \\
\label{V}
&&V = V_{\rm so} + V_{\rm tetra}, 
\end{eqnarray}
where
\begin{eqnarray}
&&
{\cal H}_{\rm cubic}
=
\sum_{\sigma=\pm}
\Bigg[ 
E_\varepsilon 
\Big(
|xy, \chi_\sigma (\phi) \rangle \langle xy, \chi_\sigma (\phi)| 
+ |yz, \chi_\sigma (\phi) \rangle \langle yz, \chi_\sigma (\phi)| 
+ |xz, \chi_\sigma (\phi) \rangle 
\langle xz, \chi_\sigma (\phi)| 
\Big) \nonumber \\
&& \hspace*{1.5cm}+ 
E_\gamma 
\Big(
|x^2-y^2, \chi_\sigma (\phi)\rangle \langle x^2-y^2, \chi_\sigma (\phi)| 
+ 
|3z^2-r^2, \chi_\sigma (\phi) \rangle \langle 3z^2-r^2, \chi_\sigma (\phi)| 
\Big) \Bigg], \\
&&V_{\rm so} = \lambda {\mbox{\boldmath $L$}} \cdot {\mbox{\boldmath $S$}}, \\
&&
V_{\rm tetra}
=\sum_{\sigma=\pm}
\Bigg[ 
\delta_\varepsilon \Big( 
|xz, \chi_\sigma (\phi) \rangle \langle xz, \chi_\sigma (\phi) | 
+ |yz, \chi_\sigma (\phi) \rangle \langle yz, \chi_\sigma (\phi) | \Big) 
\nonumber \\
&&\hspace*{1.5cm} + \delta_\gamma 
|3z^2 -r^2, \chi_\sigma (\phi) \rangle \langle 3z^2-r^2, \chi_\sigma (\phi) | 
\Bigg], 
\end{eqnarray}
and
\begin{eqnarray}
&&{\mbox{\boldmath $S$}}=(S_x, S_y, S_z), \\
&&{\mbox{\boldmath $L$}}=(L_x, L_y, L_z), \\
&&{\mbox{\boldmath $H$}}=
H \left(\cos \phi, \sin \phi, 0 \right). 
\end{eqnarray}
The above terms are explained as follows. 
The term 
${\cal H}_{\rm cubic}$ represents the crystal field of cubic symmetry. 
The term $-{\mbox{\boldmath $S$}} \cdot {\mbox{\boldmath $H$}}$ 
is the Zeeman interaction between 
the spin angular momentum ${\mbox{\boldmath $S$}}$ 
and 
the exchange field of the ferromagnet 
${\mbox{\boldmath $H$}}$, 
where 
${\mbox{\boldmath $H$}} \propto -{\mbox{\boldmath $M$}}$, 
${\mbox{\boldmath $H$}} \propto \langle {\mbox{\boldmath $S$}} \rangle$, 
and $H > 0$. 
The term 
$V_{\rm so}$ is the spin--orbit interaction, 
where $\lambda$ is the spin--orbit coupling constant 
and 
${\mbox{\boldmath $L$}}$ is the orbital angular momentum. 
The spin quantum number $S$ 
and the azimuthal quantum number $L$ 
are chosen to be 
$S=1/2$ and $L=2$.\cite{Kokado1} 
The term $V_{\rm tetra}$ is an additional term 
to reproduce the crystal field of tetragonal symmetry. 
The state $|m,\chi_\sigma (\phi) \rangle$ is expressed by 
$|m,\chi_\sigma (\phi) \rangle=|m \rangle |\chi_\sigma (\phi) \rangle$. 
The state $|m \rangle$ is the orbital state, 
defined by 
$|xy \rangle = xyf(r)$, 
$|yz \rangle = yzf(r)$, 
$|xz \rangle = xzf(r)$, 
$|x^2-y^2 \rangle = (1/2)(x^2-y^2)f(r)$, and 
$|3z^3-r^2 \rangle = [1/(2\sqrt{3})](3z^2-r^2)f(r)$, 
with $r=\sqrt{x^2 + y^2 + z^2}$ 
and $f(r) = \Gamma e^{-\zeta r}$, 
where 
$f(r)$ is the radial part of the 3d orbital, and 
$\Gamma$ and $\zeta$ are constants. 
The states 
$|xy \rangle$, $|yz \rangle$, and $|xz \rangle$ 
are 
called $d\varepsilon$ orbitals 
and 
$|x^2-y^2 \rangle$ and $|3z^2-r^2 \rangle$ are 
$d\gamma$ orbitals. 
The quantity 
$E_\varepsilon$ is the energy level of $|xy \rangle$, and 
$E_\gamma$ is that of $|x^2-y^2 \rangle$. 
The quantity 
$\Delta$ is defined as $\Delta$=$E_\gamma - E_\varepsilon$, 
$\delta_{\varepsilon}$ is the energy difference 
between $|xz \rangle$ (or $|yz \rangle$) and $|xy \rangle$, 
and $\delta_{\gamma}$ is that 
between $|3z^2-r^2 \rangle$ and $|x^2-y^2 \rangle$ 
(see Fig. \ref{energy}). 
The state 
$|\chi_\sigma (\phi) \rangle$ ($\sigma=+$ and $-$) is the spin state, i.e., 
\begin{eqnarray}
\label{+spin}
&&|\chi_+ (\phi) \rangle = \frac{1}{\sqrt{2}} \left( e^{-i\phi} |\uparrow \rangle + |\downarrow \rangle \right), \\
\label{-spin}
&&|\chi_- (\phi) \rangle = \frac{1}{\sqrt{2}} \left( -e^{-i\phi} |\uparrow \rangle + |\downarrow \rangle \right), 
\end{eqnarray}
which are eigenstates 
of $- {\mbox{\boldmath $S$}} \cdot {\mbox{\boldmath $H$}}$. 
Here, $|\chi_+ (\phi) \rangle$ ($|\chi_- (\phi) \rangle$) 
denotes the up spin state (down spin state) 
for the case in which 
the quantization axis is chosen 
along the $\langle {\mbox{\boldmath $S$}} \rangle$ direction. 
The state 
$|\uparrow \rangle$ ($|\downarrow \rangle$) represents 
the up spin state (down spin state) 
for the case in which 
the quantization axis is chosen along 
the $z$-axis. 
For ${\cal H}$ of Eq. (\ref{Hamiltonian}), 
we also assume 
the relation of 
parameters for typical ferromagnets, i.e., 
$\Delta/H \ll 1$\cite{Kokado3}, 
$|\lambda|/\Delta \ll 1$, 
$\delta_\varepsilon/\Delta \ll 1$, 
and $\delta_\gamma/\Delta \ll 1$. 

On the basis of the relation of the parameters, 
we consider ${\cal H}_0$ of Eq. (\ref{Hamiltonian0}) 
and $V$ of Eq. (\ref{V}) as 
the unperturbed term and 
the perturbed term, respectively. 
When the matrix of ${\cal H}$ of Eq. (\ref{Hamiltonian}) 
is represented in the basis set 
$|xy,\chi_\pm (\phi) \rangle$, 
$|yz,\chi_\pm (\phi) \rangle$, 
$|xz,\chi_\pm (\phi) \rangle$, 
$|x^2-y^2,\chi_\pm (\phi) \rangle$, 
and $|3z^2-r^2,\chi_\pm (\phi) \rangle$, 
the unperturbed system is degenerate 
(see Table A$\cdot$I in Ref. \citen{Kokado3}). 
We therefore use the perturbation theory 
for the case in which 
the unperturbed system is degenerate.\cite{Sakurai,Motizuki} 
As a result, we choose the following basis set 
for the subspace with 
$|xy,\chi_\pm (\phi) \rangle$, 
$|yz,\chi_\pm (\phi) \rangle$, and 
$|xz,\chi_\pm (\phi) \rangle$:\cite{Kokado3} 
\begin{eqnarray}
\label{+,+}
&&|\xi_+,\chi_+ (\phi)\rangle = A \Big[ (\delta_\varepsilon - \sqrt{\delta_\varepsilon^2 + \lambda^2} ) |xy, \chi_+ (\phi) \rangle 
+ i \lambda \sin \phi |yz, \chi_+ (\phi) \rangle 
- i\lambda \cos \phi |xz, \chi_+ (\phi) \rangle  \Big], \nonumber \\\\
\label{d,+}
&&|\delta_\varepsilon,\chi_+ (\phi) \rangle = \cos \phi |yz,\chi_+ (\phi) \rangle + \sin \phi |xz, \chi_+ (\phi) \rangle, \\
\label{-,+}
&&|\xi_-,\chi_+ (\phi) \rangle = B \Big[ (\delta_\varepsilon + \sqrt{\delta_\varepsilon^2 + \lambda^2} ) |xy, \chi_+ (\phi) \rangle 
+ i \lambda \sin \phi |yz, \chi_+ (\phi) \rangle 
- i \lambda \cos \phi |xz, \chi_+ (\phi) \rangle  \Big], 
\nonumber \\\\
\label{+,-}
&&|\xi_+,\chi_- (\phi)\rangle = A \Big[ (\delta_\varepsilon - \sqrt{\delta_\varepsilon^2 + \lambda^2} ) |xy, \chi_- (\phi) \rangle 
- i \lambda \sin \phi |yz, \chi_- (\phi) \rangle 
+ i\lambda \cos \phi |xz, \chi_- (\phi) \rangle  \Big], 
\nonumber \\\\
\label{d,-}
&&|\delta_\varepsilon,\chi_- (\phi)\rangle = \cos \phi |yz,\chi_- (\phi) \rangle + \sin \phi |xz, \chi_- (\phi) \rangle, \\
\label{-,-}
&&|\xi_-,\chi_- (\phi) \rangle = B \Big[ (\delta_\varepsilon + \sqrt{\delta_\varepsilon^2 + \lambda^2} ) |xy, \chi_- (\phi) \rangle 
- i \lambda \sin \phi |yz, \chi_- (\phi) \rangle 
+ i \lambda \cos \phi |xz, \chi_- (\phi) \rangle  \Big], \nonumber \\
\end{eqnarray}
with
\begin{eqnarray}
\label{xi_pm}
&&\xi_\pm = \frac{1}{2} \left( 
\delta_\varepsilon \pm \sqrt{\delta_\varepsilon^2 + \lambda^2}
\right), \\
\label{AAA}
&&A = 
\left(
2 \delta_\varepsilon^2 + 2 \lambda^2 - 2 \delta_\varepsilon \sqrt{\delta_\varepsilon^2 + \lambda^2}
\right)^{-1/2}, \\
\label{BBB}
&&B=
\left(
2 \delta_\varepsilon^2 + 2 \lambda^2 + 2 \delta_\varepsilon \sqrt{\delta_\varepsilon^2 + \lambda^2}
\right)^{-1/2}. 
\end{eqnarray}
Using 
$|\xi_+,\chi_\pm (\phi)\rangle$, 
$|\delta_\varepsilon,\chi_\pm (\phi) \rangle$, 
$|\xi_-,\chi_\pm (\phi) \rangle$, 
$|x^2-y^2, \chi_\pm (\phi) \rangle$, 
and 
$|3z^2-r^2, \chi_\pm (\phi) \rangle$, 
we construct the matrix of ${\cal H}$ of Eq. (\ref{Hamiltonian}) 
as shown in Table I in Ref. \citen{Kokado3}.

\subsection{Localized d states}
\label{sec_d}
Applying the first- and second-order perturbation theory 
to 
${\cal H}$ in Table I in Ref. \citen{Kokado3}, 
we obtain 
the localized d state 
$|m,\chi_\varsigma (\phi) )$, 
with 
$m=\xi_+$, $\delta_\varepsilon$, $\xi_-$, 
$x^2-y^2$, and $3z^2-r^2$ and $\varsigma=+$ and $-$, 
where $m$ ($\varsigma$) 
denotes the orbital index 
(spin index) of the dominant state in $|m,\chi_\varsigma (\phi))$ 
[see Eq. (\ref{wf_general})]. 
In general, 
$|m,\chi_\varsigma (\phi) )$ 
is written as
\begin{eqnarray}
\label{|m,chi_s)}
|m,\chi_\varsigma (\phi) ) = 
\left[ 1 - c_{m,\varsigma} (\phi) \right]
|m,\chi_\varsigma (\phi) \rangle 
+ 
\sum_{n~(\ne m)} 
\sum_{\sigma=+,-}
c_{n,\sigma}^{m,\varsigma}(\phi) 
|n,\chi_\sigma (\phi) \rangle,
\end{eqnarray}
with 
$c_{m,\varsigma} (\phi) >0$. 
Here, 
$|m,\chi_\varsigma (\phi) \rangle$ is the dominant state and 
$|n,\chi_\sigma (\phi) \rangle$ is the slightly hybridized state 
due to $V_{\rm so}$. 
The coefficient 
$1-c_{m,\varsigma}(\phi)$ 
[$c_{n,\sigma}^{m,\varsigma}(\phi)$] 
represents 
the probability amplitude of 
$|m,\chi_\varsigma (\phi) \rangle$ 
[$|n,\chi_\sigma (\phi) \rangle$], 
where 
$-c_{m,\varsigma}(\phi)$ 
means the reduction of the probability amplitude 
of $|m,\chi_\varsigma (\phi) \rangle$. 
In simple terms, 
$c_{m,\varsigma}(\phi)$ and $c_{n,\sigma}^{m,\varsigma}(\phi)$ 
represent the change in the d state 
due to $V_{\rm so}$, 
where 
$c_{m,\varsigma}(\phi)$ and $c_{n,\sigma}^{m,\varsigma}(\phi)$ 
are 0 at $\lambda=0$. 
Note that $|m,\chi_\varsigma (\phi) )$ is expressed 
up to the second order of 
$\lambda/H$, $\lambda/\Delta$, 
$\lambda/(H \pm \Delta)$, 
$\delta_t/H$, $\delta_t/\Delta$, 
and $\delta_t/(H \pm \Delta)$, 
with $t=\varepsilon$ or $\gamma$.

\subsection{Resistivity}
\label{sec_resistivity}
Using Eq. (\ref{|m,chi_s)}), 
we can obtain an expression for $\rho^i (\phi)$. 
The resistivity 
$\rho^i (\phi)$ is described 
by the two-current model,\cite{Campbell} i.e.,
\begin{eqnarray}
\label{rho^i(phi)}
&&\rho^i (\phi) 
= \frac{ \rho_+^i (\phi) \rho_-^i (\phi)}
{\rho_+^i (\phi) + \rho_-^i (\phi)}. 
\end{eqnarray}
The quantity 
$\rho_\sigma^i (\phi)$ is the resistivity of the $\sigma$ spin at $\phi$ 
in the case of $i$, 
where 
$\sigma=+$ ($-$) denotes the up spin (down spin) 
for the case in which 
the quantization axis is chosen along the direction of 
$\langle {\mbox{\boldmath $S$}} \rangle$ 
[see Eqs. (\ref{+spin}) and (\ref{-spin})]. 
The resistivity $\rho_\sigma^i (\phi)$ is written as 
\begin{eqnarray}
\label{rho_sigma^i}
&&\rho_\sigma^i (\phi)=
\frac{m_{\sigma}^*}{n_{\sigma} e^2 \tau_{\sigma}^i(\phi)}, 
\end{eqnarray}
where $e$ is the electric charge and 
$n_\sigma$ ($m^*_\sigma$) is the number density (effective mass)
of the electrons in the conduction band 
of the $\sigma$ spin.\cite{Ibach,Grosso} 
The conduction band consists of 
the s, p, and conductive d states.\cite{Kokado1} 
In addition, $1/\tau_{\sigma}^i(\phi)$ is 
the scattering rate of the conduction electron of the $\sigma$ spin 
in the case of $i$, 
expressed as
\begin{eqnarray}
\label{tau_inv}
&&\frac{1}{\tau_{\sigma}^i(\phi)} 
= \frac{1}{\tau_{s,\sigma}} + 
\sum_{m} \sum_{\varsigma=+,-}
\frac{1}{\tau_{s,\sigma \to m,\varsigma}^i (\phi)}, 
\end{eqnarray}
with
\begin{eqnarray}
\label{tau_sd_inv}
\frac{1}{\tau_{s,\sigma \to m,\varsigma}^i(\phi)} = 
\frac{2 \pi}{\hbar} n_{\rm imp} N_{\rm n} {V_{\rm imp}(R_{\rm n})}^2 
\left| (m,\chi_\varsigma (\phi)|
e^{i{\mbox{\boldmath $k$}}_\sigma^i 
\cdot {\mbox{\boldmath $r$}}},\chi_\sigma (\phi) \rangle \right|^2 
D_{m,\varsigma}^{(d)}. 
\end{eqnarray}
Here, 
$1/\tau_{s,\sigma}$ is the $s$--$s$ scattering rate, 
which is considered to be independent of $i$. 
The $s$--$s$ scattering means that 
the conduction electron of the $\sigma$ spin 
is scattered into the conduction state of the $\sigma$ spin 
by nonmagnetic impurities and phonons. 
The quantity $1/\tau_{s,\sigma \to m,\varsigma}^i(\phi)$ 
is the $s$--$d$ scattering rate 
in the case of $i$.\cite{Kokado1,Kokado2} 
The $s$--$d$ scattering means that 
the conduction electron of the $\sigma$ spin 
is scattered into 
the $\sigma$ spin state in 
$|m,\chi_\varsigma (\phi))$ of Eq. (\ref{|m,chi_s)}) 
by nonmagnetic impurities. 
The quantity $D_{m,\varsigma}^{(d)}$ represents 
the partial density of states (PDOS) of 
the wave function of the tight-binding model 
for the d state of the $m$ orbital and $\varsigma$ spin 
at the Fermi energy $E_{\mbox{\tiny F}}$, 
as described in Appendix B in Ref. \citen{Kokado1}. 
The conduction state of the $\sigma$ spin 
$|e^{i{\mbox{\boldmath $k$}}_\sigma^i 
\cdot {\mbox{\boldmath $r$}}},\chi_\sigma (\phi) \rangle$ is represented by 
the plane wave, i.e., 
\begin{eqnarray}
|e^{i{\mbox{\boldmath $k$}}_\sigma^i 
\cdot {\mbox{\boldmath $r$}}},\chi_\sigma (\phi) \rangle=
\frac{1}{\sqrt{\Omega}} e^{i{\mbox{\boldmath $k$}}_\sigma^i 
\cdot {\mbox{\boldmath $r$}}}\chi_\sigma (\phi), 
\end{eqnarray}
where 
${\mbox{\boldmath $k$}}_\sigma^i$ 
[=$(k_{x,\sigma}^i, k_{y,\sigma}^i, k_{z,\sigma}^i)$] 
is the Fermi wave vector of the $\sigma$ spin 
in the $i$ direction, 
${\mbox{\boldmath $r$}}$ is the position of the conduction electron, 
and $\Omega$ is the volume of the system. 
The quantity 
$V_{\rm imp}(R_{\rm n})$ is 
the scattering potential at $R_{\rm n}$ 
due to a single impurity, 
where $R_{\rm n}$ is the distance between 
the impurity and the nearest-neighbor host atom.\cite{Kokado1} 
The quantity 
$N_{\rm n}$ is the number of nearest-neighbor host atoms 
around a single impurity,\cite{Kokado1} 
$n_{\rm imp}$ is the number density of impurities, 
and 
$\hbar$ is the Planck constant $h$ divided by 2$\pi$.

We calculate 
the overlap integral 
$\langle m,\chi_\sigma (\phi)|
e^{i{\mbox{\boldmath $k$}}_\sigma^i 
\cdot {\mbox{\boldmath $r$}}},\chi_\sigma (\phi) \rangle$ 
in Eq. (\ref{tau_sd_inv}) 
using Eq. (C$\cdot$1) in Ref. \citen{Kokado3}. 
The overlap integrals of 
${\mbox{\boldmath $I$}}//[100]$, 
${\mbox{\boldmath $I$}}//[110]$, 
and ${\mbox{\boldmath $I$}}//[001]$ are as follows: 

\begin{enumerate}
\item[(i)] ${\mbox{\boldmath $I$}}//[100]$ \\
In the case of 
${\mbox{\boldmath $I$}}//[100]$ 
corresponding to ${\mbox{\boldmath $k$}}_\sigma^{[100]}=(k_\sigma, 0, 0)$, 
the overlap integral becomes 
\begin{eqnarray}
\label{s100_xy}
&&\langle xy,\chi_{\sigma'} (\phi) |e^{ik_\sigma x},\chi_\sigma (\phi) \rangle=\langle yz,\chi_{\sigma'} (\phi) |e^{ik_\sigma x},\chi_\sigma (\phi) \rangle 
=\langle xz,\chi_{\sigma'} (\phi) |e^{ik_\sigma x},\chi_\sigma (\phi) \rangle=0,\nonumber \\\\
\label{s100_x2-y2}
&&\langle x^2 -y^2,\chi_{\sigma'} (\phi)|e^{ik_\sigma x},\chi_\sigma (\phi) \rangle=
\frac{1}{2} g_\sigma \delta_{\sigma,\sigma'}, \\
\label{s100_3z2-r2}
&&\langle 3z^2 -r^2,\chi_{\sigma'} (\phi)|e^{ik_\sigma x},\chi_\sigma (\phi) \rangle=
-\frac{1}{2\sqrt{3}} g_\sigma \delta_{\sigma,\sigma'}, 
\end{eqnarray}
with 
\begin{eqnarray}
\label{g_sigma}
&&g_\sigma = 
- \frac{192 \pi \Gamma \zeta k_\sigma^2 
}{\sqrt{\Omega}(k_\sigma^2 + \zeta^2)^4}.
\end{eqnarray}
The scatterings 
from 
the plane wave to $|3z^2-r^2, \chi_\sigma (\phi) \rangle$ 
and $|x^2 -y^2,\chi_{\sigma} (\phi) \rangle$ 
are thus allowed. 
Using Eqs. (\ref{s100_xy}) and (\ref{+,+})$-$(\ref{-,-}), we also have 
\begin{eqnarray}
\label{s100_+}
\langle \xi_+,\chi_{\sigma'} (\phi) |e^{ik_\sigma x},\chi_\sigma (\phi) \rangle=\langle \delta_\varepsilon,\chi_{\sigma'} (\phi) |e^{ik_\sigma x},\chi_\sigma (\phi) \rangle 
=\langle \xi_-,\chi_{\sigma'} (\phi) |e^{ik_\sigma x},\chi_\sigma (\phi) \rangle=0. 
\end{eqnarray}

\item[(ii)] ${\mbox{\boldmath $I$}}//[110]$ \\
In the case of 
${\mbox{\boldmath $I$}}//[110]$ 
corresponding to 
${\mbox{\boldmath $k$}}_\sigma^{[110]}=(k_\sigma, k_\sigma, 0)/\sqrt{2}$, 
the overlap integral is
\begin{eqnarray}
\label{s110_xy}
&&\langle xy,\chi_{\sigma'} (\phi)|e^{i(k_\sigma x + k_\sigma y)/\sqrt{2}},\chi_\sigma (\phi) \rangle=
\frac{1}{2}g_\sigma \delta_{\sigma,\sigma'}, \\ 
\label{s110_yz}
&&
\langle yz,\chi_{\sigma'} (\phi) |e^{i(k_\sigma x + k_\sigma y)/\sqrt{2}},\chi_\sigma (\phi) \rangle 
=\langle xz,\chi_{\sigma'} (\phi) |e^{i(k_\sigma x + k_\sigma y)/\sqrt{2}},\chi_\sigma (\phi) \rangle = 0, \nonumber \\\\
\label{s110_x2-y2}
&&
\langle x^2-y^2,\chi_{\sigma'} (\phi) |e^{i(k_\sigma x + k_\sigma y)/\sqrt{2}},\chi_\sigma (\phi) \rangle=0, \\
\label{s110_3z2-r2}
&&\langle 3z^2 -r^2,\chi_{\sigma'} (\phi)|e^{i(k_\sigma x + k_\sigma y)/\sqrt{2}},\chi_\sigma (\phi) \rangle=
-\frac{1}{2\sqrt{3}} g_\sigma \delta_{\sigma,\sigma'}. 
\end{eqnarray}
The scatterings 
from 
the plane wave to 
$|xy, \chi_\sigma (\phi) \rangle$ and 
$|3z^2-r^2, \chi_\sigma (\phi) \rangle$ 
are thus allowed. 
Using Eqs. (\ref{s110_xy}), (\ref{s110_yz}), and (\ref{+,+})$-$(\ref{-,-}), 
we also have
\begin{eqnarray}
\label{s110_+}
&&\langle \xi_+,\chi_{\sigma'} (\phi)|e^{i(k_\sigma x + k_\sigma y)/\sqrt{2}},\chi_\sigma (\phi) \rangle=
\frac{1}{2}A
\left( \delta_e - \sqrt{\delta_e^2 + \lambda^2} \right) 
g_\sigma \delta_{\sigma,\sigma'}, \\ 
\label{s110_d}
&&\langle \delta_\varepsilon,\chi_{\sigma'} (\phi)|e^{i(k_\sigma x + k_\sigma y)/\sqrt{2}},\chi_\sigma (\phi) \rangle=0, \\
\label{s110_-}
&&\langle \xi_-,\chi_{\sigma'} (\phi)|e^{i(k_\sigma x + k_\sigma y)/\sqrt{2}},\chi_\sigma (\phi) \rangle=
\frac{1}{2} B 
\left( \delta_e + \sqrt{\delta_e^2 + \lambda^2} \right) 
g_\sigma \delta_{\sigma,\sigma'}, 
\end{eqnarray}
where $A$ and $B$ have been given by Eqs. (\ref{AAA}) and (\ref{BBB}), 
respectively.

\item[(iii)] ${\mbox{\boldmath $I$}}//[001]$ \\
In the case of 
${\mbox{\boldmath $I$}}//[001]$ 
corresponding to 
${\mbox{\boldmath $k$}}_\sigma^{[001]}=(0, 0, k_\sigma)$, 
the overlap integral is
\begin{eqnarray}
\label{s001_xy}
&&\langle xy,\chi_{\sigma'} (\phi) |e^{ik_\sigma z},\chi_\sigma (\phi) \rangle=\langle yz,\chi_{\sigma'} (\phi) |e^{ik_\sigma z},\chi_\sigma (\phi) \rangle 
=\langle xz,\chi_{\sigma'} (\phi) |e^{ik_\sigma z},\chi_\sigma (\phi) \rangle=0,\nonumber \\\\
\label{s001_x2-y2}
&&\langle x^2-y^2,\chi_{\sigma'} (\phi) |e^{ik_\sigma z},\chi_\sigma (\phi) \rangle=0, \\ 
\label{s001_3z2-r2}
&&\langle 3z^2 -r^2,\chi_{\sigma'} (\phi)|e^{i k_\sigma z},\chi_\sigma (\phi) \rangle=
\frac{1}{\sqrt{3}} g_\sigma \delta_{\sigma,\sigma'}. 
\end{eqnarray}
Only the scattering from 
the plane wave to $|3z^2-r^2, \chi_\sigma (\phi) \rangle$ 
is thus allowed. 
Using Eqs. (\ref{s001_xy}) and (\ref{+,+})$-$(\ref{-,-}), we also have
\begin{eqnarray}
\label{s001_+}
\langle \xi_+,\chi_{\sigma'} (\phi) |e^{ik_\sigma z},\chi_\sigma (\phi) \rangle=\langle \delta_\varepsilon,\chi_{\sigma'} (\phi) |e^{ik_\sigma z},\chi_\sigma (\phi) \rangle 
=\langle \xi_-,\chi_{\sigma'} (\phi) |e^{ik_\sigma z},\chi_\sigma (\phi) \rangle=0. 
\end{eqnarray}
\end{enumerate}

Substituting the above results into Eq. (\ref{tau_sd_inv}), 
we obtain the expression for $\rho_\sigma^i(\phi)$ of Eq. (\ref{rho_sigma^i}) 
as shown in 
Appendix \ref{appen_resis}. 
Here, $\rho_\sigma^i(\phi)$ 
is expressed by using 
the following quantities\cite{Kokado2}: 
\begin{eqnarray}
\label{rho_s_pm}
&&\rho_{s,\sigma}= \frac{m_\sigma^*}{n_\sigma e^2 \tau_{s,\sigma}}, \\
\label{rho_m_pm}
&&\rho_{s,\sigma \to m,\varsigma}
= \frac{m_\sigma^*}{n_\sigma e^2 \tau_{s,\sigma \to m,\varsigma}}, 
\end{eqnarray}
where $\rho_{s,\sigma}$ is the $s$--$s$ resistivity 
and 
$\rho_{s,\sigma \to m,\varsigma}$ is the $s$--$d$ resistivity. 
The $s$--$d$ scattering rate $1/\tau_{s,\sigma \to m,\varsigma}$ 
is defined by
\begin{eqnarray}
\label{1/tau_m_pm}
&&\frac{1}{\tau_{s,\sigma \to m,\varsigma}}=
\frac{2\pi}{\hbar} n_{\rm imp}N_{\rm n} {V_{\rm imp}(R_{\rm n})}^2  
\left|\langle 3z^2 -r^2,\chi_\sigma (\phi) | e^{ik_\sigma z}, \chi_\sigma (\phi) \rangle \right|^2 
D_{m,\varsigma}^{(d)} \nonumber \\
&&\hspace*{1.5cm}=
\frac{2\pi}{\hbar} 
n_{\rm imp}N_{\rm n}\frac{1}{3}v_\sigma^2 D_{m, \varsigma}^{(d)},
\end{eqnarray}
with 
\begin{eqnarray}
\label{v_sigma}
v_{\sigma}=V_{\rm imp}(R_{\rm n}) g_\sigma, 
\end{eqnarray}
where 
$g_\sigma$ is given by Eq. (\ref{g_sigma}). 
The overlap integral 
$\langle 3z^2 -r^2,\chi_\sigma (\phi) | 
e^{ik_\sigma z}, \chi_\sigma (\phi) \rangle$ 
in Eq. (\ref{1/tau_m_pm}) 
can be calculated by using 
Eq. (C$\cdot$1) in Ref. \citen{Kokado3}. 
Here, 
Eq. (\ref{1/tau_m_pm}) 
has been introduced 
to investigate the relation 
between the present result and 
the previous ones\cite{Campbell,Kokado1,Kokado2} 
(also see Appendix \ref{appen_CFJ}).

We also note that, 
as found from Eq. (\ref{1/tau_m_pm}), 
$\rho_{s,\sigma \to m,\varsigma}$ of Eq .(\ref{rho_m_pm}) satisfies 
\begin{eqnarray}
\label{proportional}
\rho_{s,\sigma \to m,\varsigma} \propto D_{m,\varsigma}^{(d)}. 
\end{eqnarray}
This relation is useful to give a physical explanation for $C_j^i$.

\section{Application}
\label{sec_calc}
We apply the theory of Sect. \ref{sec_theory} to ferromagnets 
with $D_{m,+}^{(d)}=0$ and $D_{m,-}^{(d)} \ne 0$. 
Using $\rho_\sigma^i(\phi_i)$ in Appendix \ref{appen_resis}, 
we obtain 
AMR$^i(\phi_i)$ of Eq. (\ref{AMR^i(phi_i)}) for the ferromagnets. 
The AMR ratio AMR$^i(\phi_i)$ is expressed 
up to the second order of 
$\lambda/H$, $\lambda/\Delta$, 
$\lambda/(H \pm \Delta)$, 
$\delta_t/H$, $\delta_t/\Delta$, 
and $\delta_t/(H \pm \Delta)$, 
with $t=\varepsilon$ or $\gamma$. 
Here, we introduce 
\begin{eqnarray}
\label{r_s-s}
&& r=\frac{\rho_{s,-}}{\rho_{s,+}}, \\
\label{r_s-d}
&& r_{s,\sigma \to m,-}=
\frac{\rho_{s,\sigma \to m,-}}{\rho_{s,+}}, 
\end{eqnarray}
in accordance with our previous study\cite{Kokado2}. 
In addition, we set 
\begin{eqnarray}
\label{epsilon1}
&&r_{s, \sigma \to \delta_\varepsilon,-}
\equiv r_{s, \sigma \to \varepsilon 1,-}, \\
\label{epsilon2}
&&r_{s, \sigma \to \xi_+,-}= r_{s, \sigma \to \xi_-,-}
 \equiv r_{s, \sigma \to \varepsilon 2,-},  
\end{eqnarray}
for simplicity.

\subsection{${\mbox{\boldmath $I$}}//[100]$} 
\label{sub_I//100}
Using 
Eqs. (\ref{AMR^i(phi_i)}), (\ref{rho^i(phi)}), and (\ref{rho_pm^100}), 
we obtain 
${\rm AMR}^{[100]}(\phi)$: 
\begin{eqnarray}
\label{AMR^100}
{\rm AMR}^{[100]}(\phi)
= C_0^{[100]} + C_2^{[100]} \cos 2\phi + C_4^{[100]} \cos 4\phi. 
\end{eqnarray}
Here, $C_0^{[100]}$ is determined 
so as to satisfy ${\rm AMR}^{[100]}(\pi/2)=0$. 
In addition, 
$C_2^{[100]}$ and $C_4^{[100]}$ are expressed as 
Eqs. (\ref{C_2^100_rho}) and (\ref{C_4^100_rho}), respectively. 
Using Eqs. (\ref{C_2^100_rho}) and (\ref{C_4^100_rho}), 
Eq. (43) in Ref. \citen{Kokado3}, 
Eq. (45) in Ref. \citen{Kokado3}, 
Eq. (46) in Ref. \citen{Kokado3}, 
Eq. (2) in Ref. \citen{Kokado3_1}, 
and 
Eq. (3) in Ref. \citen{Kokado3_1}, 
we derive expressions for $C_2^{[100]}$ and $C_4^{[100]}$, 
where 
$\rho_{s,\sigma \to m,+}=0$ due to $D_{m,+}^{(d)}=0$ is taken into account. 
The respective expressions are 
given in Sect. \ref{appen_C_j^100}. 

\subsection{${\mbox{\boldmath $I$}}//[110]$} 
\label{sub_I//110}
Using 
Eqs. (\ref{AMR^i(phi_i)}), (\ref{rho^i(phi)}), and (\ref{rho_pm^110}), 
we obtain ${\rm AMR}^{[110]}(\phi')$: 
\begin{eqnarray}
\label{AMR^110}
{\rm AMR}^{[110]}(\phi')
= C_0^{[110]} + C_2^{[110]} \cos 2\phi' + C_4^{[110]} \cos 4\phi'
+ C_6^{[110]} \cos 6\phi' + C_8^{[110]} \cos 8\phi'. 
\end{eqnarray}
Here, $C_0^{[110]}$ 
is determined 
so as to satisfy ${\rm AMR}^{[110]}(\pi/2)=0$. 
In addition, 
$C_2^{[110]}$, $C_4^{[110]}$, $C_6^{[110]}$, and $C_8^{[110]}$ 
are expressed as 
Eqs. (\ref{C_2^110_rho}), (\ref{C_4^110_rho}), (\ref{C_6^110_rho}), 
and (\ref{C_8^110_rho}), respectively. 
Using 
Eqs. 
(\ref{C_2^110_rho})$-$(\ref{C_8^110_rho}), 
(\ref{rho_0+^110(0)}), (\ref{rho_0-^110(0)}), 
and (\ref{rho_2+^110(2)})$-$(\ref{rho_8-^110(2)}), 
we derive expressions for $C_2^{[110]}$, $C_4^{[110]}$, 
$C_6^{[110]}$, and $C_8^{[110]}$. 
The respective expressions are 
given in Sect. \ref{appen_C_j^110}. 

\subsection{${\mbox{\boldmath $I$}}//[001]$} 
\label{sub_I//001}
Using 
Eqs. (\ref{AMR^i(phi_i)}), (\ref{rho^i(phi)}), and (\ref{rho_pm^001}), 
we obtain ${\rm AMR}^{[001]}(\phi)$: 
\begin{eqnarray}
\label{AMR^001}
{\rm AMR}^{[001]}(\phi)
= C_0^{[001]} + C_4^{[001]} \cos 4\phi. 
\end{eqnarray}
Here, $C_0^{[001]}$ 
is determined 
so as to satisfy ${\rm AMR}^{[001]}(\pi/2)=0$. 
In addition, 
$C_4^{[001]}$ 
is expressed as 
Eq. (\ref{C_4^001_rho}). 
Using 
Eqs. (\ref{C_4^001_rho}), (\ref{rho_0+^001(0)}), 
(\ref{rho_0-^001(0)}), (\ref{rho_4+^001(2)}), and (\ref{rho_4-^001(2)}), 
we derive an expression for $C_4^{[001]}$. 
The respective expressions are
given in Sect. \ref{appen_C_j^001}. 
Note that the feature 
that the $\phi$-dependent term is only the $\cos 4\phi$ term 
is also found in the expression by 
D$\ddot{\rm o}$ring,\cite{Doring} i.e., Eq. (\ref{Doring_001}).

\subsection{Simplified system} 
\label{simplified}
On the basis of the above-mentioned $C_j^i$, 
we obtain a simple expression for $C_j^i$ for the simplified system. 
In this system, 
we assume 
\begin{eqnarray}
\label{gamma}
r_{s,- \to x^2-y^2, -} = r_{s,- \to 3z^2-r^2, -} \equiv  
r_{s,- \to \gamma, -}, 
\end{eqnarray}
which corresponds to 
$\rho_{s,- \to x^2-y^2, -} = \rho_{s,- \to 3z^2-r^2, -}$, i.e., 
$D_{x^2-y^2,-}^{(d)}=D_{3z^2-r^2,-}^{(d)}$ 
[see Eqs. (\ref{r_s-d}) and (\ref{proportional})]. 
This assumption 
may be valid for the system of 
$D_{x^2-y^2,-}^{(d)} \sim D_{3z^2-r^2,-}^{(d)}$ 
and/or $|\lambda|/\delta_\gamma \ll 1$. 
The reason is that 
the terms with 
$r_{s,- \to 3z^2-r^2, -}-r_{s,- \to x^2-y^2, -}$ 
in $C_j^i$ have 
$\left( \frac{\lambda}{\delta_\gamma} \right) 
\left( r_{s,- \to 3z^2-r^2, -} - r_{s,- \to x^2-y^2, -} \right)$, 
$\left( \frac{\lambda}{\delta_\gamma} \right)^2 
\left( r_{s,- \to 3z^2-r^2, -} - r_{s,- \to x^2-y^2, -} \right)$, 
and 
$\left( \frac{\lambda}{\delta_\gamma} \right)^2 
\left( r_{s,- \to 3z^2-r^2, -} - r_{s,- \to x^2-y^2, -} \right)^2$. 
We also use 
$(\frac{\lambda}{H \pm \Delta})^2 \approx \left( \frac{\lambda}{H} \right)^2$, 
$\frac{\lambda^2}{\Delta (H \pm \Delta)} \approx \frac{\lambda^2}{H\Delta}
\mp \left( \frac{\lambda}{H} \right)^2$, 
and 
$\frac{\lambda^2}{H (H \pm \Delta)} \approx 
\left( \frac{\lambda}{H} \right)^2$ 
due to $\Delta/H \ll 1$.

For this system,  we consider three types:
\begin{enumerate}
\item[(i)] type A: 
generalized strong ferromagnet with 
the $s$--$d$ scattering ``$s,+ \to d,-$ and $s,- \to d,-$'', 
\item[(ii)] type B: 
half-metallic ferromagnet 
with the dominant $s$--$d$ scattering ``$s,- \to d,-$'', and 
\item[(iii)] type C: 
specified strong ferromagnet 
with the dominant $s$--$d$ scattering ``$s,+ \to d,-$''. 
\end{enumerate}
In Tables \ref{table1}, \ref{table2}, and \ref{table3}, 
we show 
$C_j^i$ for types A, B, and C, respectively. 
The coefficient $C_j^i$ for type A is derived 
by imposing Eq. (\ref{gamma}) 
on Eqs. (\ref{C_0^100})$-$(\ref{C_4^100}), 
(\ref{C_0^110})$-$(\ref{C_8^110}), (\ref{C_0^001}), and (\ref{C_4^001}). 
The coefficient $C_j^i$ for type B is obtained 
by imposing 
$r \ll 1$, $r_{s,\sigma \to \varepsilon 1,- } \ll 1$, $r_{s,\sigma \to \varepsilon 2,- } \ll 1$, and $r_{s,\sigma \to \gamma,- } \ll 1$ 
on $C_j^i$ for type A in Table \ref{table1}. 
The coefficient $C_j^i$ for type C is obtained 
by imposing 
$r \gg r_{s,\sigma \to \varepsilon 1,- }$, $r \gg r_{s,\sigma \to \varepsilon 2,- }$, $r \gg r_{s,\sigma \to \gamma,- }$, and $r \gg 1$ 
on $C_j^i$ for type A in Table \ref{table1}, 
where 
$r$ is set to be large enough 
for the term including $r$ in the numerator 
to become dominant in each $C_j^i$ in spite of $\Delta/H \ll 1$. 
Note here that 
$C_6^{[110]}$ and $C_8^{[110]}$ are regarded as 0, 
because $C_6^{[110]}$ and $C_8^{[110]}$ 
in Table \ref{table1} 
include $r$ only in the respective denominators 
and then they become smaller than the other $C_j^i$.

We also mention that 
$C_2^{[100]}$ and $C_4^{[100]}$ for type A 
in Table \ref{table1} 
are, respectively, the coefficients in our previous study, i.e., 
Eqs. (61) and (62) in Ref. \citen{Kokado3}, 
where 
$(\frac{\lambda}{H \pm \Delta})^2 \approx (\frac{\lambda}{H})^2$ 
is used in this study. 
In addition, 
AMR$^{[100]}(0)$ of Eq. (\ref{AMR^100}) 
with $C_2^{[100]}$ in Table \ref{table1} 
and AMR$^{[110]}(0)$ of Eq. (\ref{AMR^110}) 
with $C_2^{[110]}$ and $C_6^{[110]}$ in Table \ref{table1} 
correspond to the CFJ model\cite{Campbell} 
under the condition of the CFJ model 
(see Appendix \ref{appen_CFJ}).\cite{[001]}

\begin{table}[ht]
\caption{
The coefficient $C_j^i$ 
for type A, i.e., the generalized strong ferromagnet 
with 
the $s$--$d$ scattering ``$s,+ \to d,-$ and $s,- \to d,-$''. 
Type A has 
$D_{m,+}^{(d)}=0$, 
$D_{m,-}^{(d)}\ne 0$, 
$r_{s,\sigma \to \delta_\varepsilon,-} 
\equiv r_{s,\sigma \to \varepsilon 1, -}$, 
$r_{s,\sigma \to \xi_+,-}=r_{s,\sigma \to \xi_-,-} \equiv 
r_{s,\sigma \to \varepsilon 2, -}$, 
$r_{s,\sigma \to 3z^2-r^2, -} = r_{s,\sigma \to x^2-y^2, -} \equiv  
r_{s,\sigma \to \gamma, -}$, 
and $\Delta/H \ll 1$. 
This $C_j^i$ is obtained by imposing 
$r_{s,\sigma \to 3z^2-r^2, -} = r_{s,\sigma \to x^2-y^2, -} \equiv  
r_{s,\sigma \to \gamma, -}$ 
and $\Delta/H \ll 1$ on 
Eqs. (\ref{C_0^100})$-$(\ref{C_4^001}). 
}
\begin{center}
\begin{tabular}{l}
\hline 
Coefficient \\
\hline 
${\mbox{\boldmath $I$}}//[100]$ \\
$C_0^{[100]} = C_2^{[100]} - C_4^{[100]}$ \\
$C_2^{[100]}=
\frac{3}{8} 
\frac{1}{1 + r + r_{s,- \to \gamma,-}}
\Bigg[ 
\left( \frac{\lambda}{\Delta} \right)^2 \frac{ r_{s, - \to \gamma, - }- r_{s,- \to \varepsilon 1, -}}{r + r_{s,- \to \gamma,-}}
- 
\left( \frac{\lambda}{H} \right)^2 \frac{ r_{s, - \to \gamma, - }}{r + r_{s,- \to \gamma,-}}
+ 
\left( \frac{\lambda}{H} \right)^2 r_{s, + \to \varepsilon 2, - }(r + r_{s,- \to \gamma,-}) \Bigg]$ \\
$C_4^{[100]}=
\frac{3}{32} 
\frac{1}{1 +r + r_{s,- \to \gamma,-}}
\Bigg[ 
\left( \frac{\lambda}{\Delta} \right)^2 \frac{ r_{s, - \to \varepsilon 1, - }- r_{s,- \to \varepsilon 2, -}}{r + r_{s,- \to \gamma,-}} 
+ \left( \frac{\lambda}{H} \right)^2 
(r_{s, + \to \varepsilon 2, - }- r_{s,+ \to \varepsilon 1, -})
(r + r_{s,- \to \gamma,-}) \Bigg] $ 
\\\\[-0.3cm]
${\mbox{\boldmath $I$}}//[110]$ \\
$C_0^{[110]}=C_2^{[110]} - C_4^{[110]} + C_6^{[110]} - C_8^{[110]}$ \\
$C_2^{[110]}=
\frac{3}{8} 
\frac{1}
{1+r +  \frac{3}{4} r_{s, - \to \varepsilon 2, - } + \frac{1}{4} r_{s,- \to \gamma,-}}
\Bigg[ 
\left( \frac{\lambda}{\Delta} \right)^2 
\frac{
\frac{3}{4} r_{s, - \to \varepsilon 1, - } + \frac{1}{4} r_{s, - \to \varepsilon 2, - } - r_{s,- \to \gamma, -} }
{r +  \frac{3}{4} r_{s, - \to \varepsilon 2, - } + \frac{1}{4} r_{s,- \to \gamma,-}} $
\nonumber \\
\hspace{1.2cm} $+ \frac{\lambda^2}{H\Delta} 
\frac{
\frac{3}{4} r_{s, - \to \varepsilon 1, - } - \frac{7}{4} r_{s, - \to \varepsilon 2, - } + r_{s,- \to \gamma, -} }
{r +  \frac{3}{4} r_{s, - \to \varepsilon 2, - } + \frac{1}{4} r_{s,- \to \gamma,-}} 
+ \frac{3}{4} \left( \frac{\lambda}{H} \right)^2 
\frac{r_{s, - \to \varepsilon 1, - } - r_{s, - \to \varepsilon 2, - }}
{r +  \frac{3}{4} r_{s, - \to \varepsilon 2, - } + \frac{1}{4} r_{s,- \to \gamma,-}} $ \\
\hspace{1.2cm} 
$- \left( \frac{\lambda}{H} \right)^2 
\frac{r_{s, - \to \gamma, - }}
{r +  \frac{3}{4} r_{s, - \to \varepsilon 2, - } + \frac{1}{4} r_{s,- \to \gamma,-}} + \left( \frac{\lambda}{H} \right)^2 
r_{s, + \to \varepsilon 1, - } 
\left( r +  \frac{3}{4} r_{s, - \to \varepsilon 2, - } + \frac{1}{4} r_{s,- \to \gamma,-} \right) \Bigg]$ \\
$C_4^{[110]}= \frac{3}{32} \frac{ 1}{1 + r +  \frac{3}{4} r_{s, - \to \varepsilon 2, - } + \frac{1}{4} r_{s,- \to \gamma,-}} 
\Bigg[ \left( \frac{\lambda}{\Delta} \right)^2 
\frac{r_{s, - \to \varepsilon 2, - } - r_{s, - \to \varepsilon 1, - }}{r +  \frac{3}{4} r_{s, - \to \varepsilon 2, - }  + \frac{1}{4} r_{s,- \to \gamma,-}} $ \\
\hspace{1.2cm}  $ + \left( \frac{\lambda}{H} \right)^2 
(r_{s,+ \to \varepsilon 1, - } - r_{s,+ \to \varepsilon 2, - })
\left( r +  \frac{3}{4} r_{s, - \to \varepsilon 2, - }  + \frac{1}{4} r_{s,- \to \gamma,-} \right) 
\Bigg] $ \\
$C_6^{[110]}=\frac{9}{32} 
\left[ \left( \frac{\lambda}{\Delta} \right)^2
+ \frac{\lambda^2}{H \Delta} 
+ \left( \frac{\lambda}{H} \right)^2 \right] 
\frac{ r_{s, - \to \varepsilon 2, - } - r_{s, - \to \varepsilon 1, - }}{\left( r +  \frac{3}{4} r_{s, - \to \varepsilon 2, - } + \frac{1}{4} r_{s,- \to \gamma,-} \right) \left(1 + r +  \frac{3}{4} r_{s, - \to \varepsilon 2, - } + \frac{1}{4} r_{s,- \to \gamma,-} \right) } $ \\
$C_8^{[110]}=\frac{27}{128}
\left[ \left( \frac{\lambda}{\Delta} \right)^2
+ 2\frac{\lambda^2}{H \Delta} 
+ 3\left( \frac{\lambda}{H} \right)^2 \right] 
\frac{ r_{s, - \to \varepsilon 2, - } - r_{s, - \to \varepsilon 1, - }}{\left( r +  \frac{3}{4} r_{s, - \to \varepsilon 2, - } + \frac{1}{4} r_{s,- \to \gamma,-} \right) \left(1 + r +  \frac{3}{4} r_{s, - \to \varepsilon 2, - } + \frac{1}{4} r_{s,- \to \gamma,-} \right) } $  \\\\[-0.3cm]
${\mbox{\boldmath $I$}}//[001]$ \\
$C_0^{[001]}=-C_4^{[001]}$ \\
$C_4^{[001]}=\frac{3}{8}
\frac{1}{1 + r + r_{s,- \to \gamma, -}} 
\left[ 
\left( \frac{\lambda }{\Delta} \right)^2 
\frac{r_{s,- \to \varepsilon 1, -} - r_{s,- \to \varepsilon 2, -}}
{r + r_{s,- \to \gamma, -}}
+ \left( \frac{\lambda }{H} \right)^2 
(r_{s,+ \to \varepsilon 2, -} - r_{s,+ \to \varepsilon 1, -})
(r + r_{s,- \to \gamma, -}) \right] $ \\
\hline 
\end{tabular}
\end{center}
\label{table1}
\end{table}

\begin{table}[ht]
\caption{
The coefficient $C_j^i$ 
for type B, i.e., the half-metallic ferromagnet 
with the dominant $s$--$d$ scattering ``$s,- \to d,-$''. 
This $C_j^i$ is obtained 
by imposing 
$r \ll 1$, $r_{s,\sigma \to \varepsilon 1,- } \ll 1$, $r_{s,\sigma \to \varepsilon 2,- } \ll 1$, and $r_{s,\sigma \to \gamma,- } \ll 1$ 
on $C_j^i$ for type A in Table \ref{table1}. 
}
\begin{center}
\begin{tabular}{l}
\hline 
Coefficient \\
\hline 
${\mbox{\boldmath $I$}}//[100]$ \\
$C_0^{[100]} = C_2^{[100]} - C_4^{[100]}$ \\
$C_2^{[100]}=\displaystyle{\frac{3}{8} \left[ \left( \frac{\lambda}{\Delta} \right)^2 \frac{ r_{s, - \to \gamma, - }- r_{s,- \to \varepsilon 1, -}}{r + r_{s,- \to \gamma,-}} 
- 
\left( \frac{\lambda}{H} \right)^2 \frac{ r_{s, - \to \gamma, - }}{r + r_{s,- \to \gamma,-}} \right]}$ \\
$C_4^{[100]}=\displaystyle{\frac{3}{32} \left( \frac{\lambda}{\Delta} \right)^2 \frac{ r_{s, - \to \varepsilon 1, - }- r_{s,- \to \varepsilon 2, -}}{r + r_{s,- \to \gamma,-}}}$ 
\\\\[-0.3cm]
${\mbox{\boldmath $I$}}//[110]$ \\
$C_0^{[110]}=C_2^{[110]} - C_4^{[110]} + C_6^{[110]} - C_8^{[110]}$ \\
$C_2^{[110]}=\displaystyle{
\frac{3}{8} 
\Bigg[ 
\left( \frac{\lambda}{\Delta} \right)^2 \frac{ \frac{3}{4} r_{s, - \to \varepsilon 1, - } + \frac{1}{4} r_{s, - \to \varepsilon 2, - } - r_{s,- \to \gamma, -}}{r +  \frac{3}{4} r_{s, - \to \varepsilon 2, - } + \frac{1}{4} r_{s,- \to \gamma,-}}
+ \frac{\lambda^2}{H\Delta} 
\frac{
\frac{3}{4} r_{s, - \to \varepsilon 1, - } - \frac{7}{4} r_{s, - \to \varepsilon 2, - } + r_{s,- \to \gamma, -}}{r +  \frac{3}{4} r_{s, - \to \varepsilon 2, - } + \frac{1}{4} r_{s,- \to \gamma,-}}}$ \\
\hspace{1.2cm}
$\displaystyle{ + \frac{3}{4} \left( \frac{\lambda}{H} \right)^2 
\frac{r_{s, - \to \varepsilon 1, - } - r_{s, - \to \varepsilon 2, - }}
{r +  \frac{3}{4} r_{s, - \to \varepsilon 2, - } + \frac{1}{4} r_{s,- \to \gamma,-}}
- \left( \frac{\lambda}{H} \right)^2 
\frac{r_{s, - \to \gamma, - }}
{r +  \frac{3}{4} r_{s, - \to \varepsilon 2, - } + \frac{1}{4} r_{s,- \to \gamma,-}}
\Bigg] 
}$ \\
$C_4^{[110]}=\displaystyle{\frac{3}{32} \left( \frac{\lambda}{\Delta} \right)^2 \frac{ r_{s, - \to \varepsilon 2, - } - r_{s, - \to \varepsilon 1, - }}{r +  \frac{3}{4} r_{s, - \to \varepsilon 2, - } + \frac{1}{4} r_{s,- \to \gamma,-}}}$ \\
$C_6^{[110]}=
\displaystyle{\frac{9}{32} 
\left[ \left( \frac{\lambda}{\Delta} \right)^2
+ \frac{\lambda^2}{H \Delta} 
+ \left( \frac{\lambda}{H} \right)^2 \right] 
\frac{ r_{s, - \to \varepsilon 2, - } - r_{s, - \to \varepsilon 1, - }}{r +  \frac{3}{4} r_{s, - \to \varepsilon 2, - } + \frac{1}{4} r_{s,- \to \gamma,-}}}$ \\
$C_8^{[110]}=\displaystyle{\frac{27}{128} 
\left[ \left( \frac{\lambda}{\Delta} \right)^2
+ 2\frac{\lambda^2}{H \Delta} 
+ 3\left( \frac{\lambda}{H} \right)^2 \right] 
\frac{ r_{s, - \to \varepsilon 2, - } - r_{s, - \to \varepsilon 1, - }}{r +  \frac{3}{4} r_{s, - \to \varepsilon 2, - } + \frac{1}{4} r_{s,- \to \gamma,-}}}$ 
\\\\[-0.3cm]
${\mbox{\boldmath $I$}}//[001]$ \\
$C_0^{[001]}=-C_4^{[001]}$ \\
$C_4^{[001]}=\displaystyle{\frac{3}{8}
\left( \frac{\lambda }{\Delta} \right)^2 
\frac{r_{s,- \to \varepsilon 1, -} - r_{s,- \to \varepsilon 2, -}}
{r + r_{s,- \to \gamma, -}}}$ \\
\hline 
\end{tabular}
\end{center}
\label{table2}
\end{table}

\begin{table}[ht]
\caption{
The coefficient $C_j^i$ 
for type C, i.e., 
the specified strong ferromagnet 
with the dominant $s$--$d$ scattering ``$s,+ \to d,-$''. 
This $C_j^i$ is obtained 
by imposing 
$r \gg r_{s,\sigma \to \varepsilon 1,- }$, $r \gg r_{s,\sigma \to \varepsilon 2,- }$, $r \gg r_{s,\sigma \to \gamma,- }$, and $r \gg 1$ 
on $C_j^i$ for type A in Table \ref{table1}, 
where 
$r$ is set to be large enough 
for the term 
including $r$ in the numerator 
to become dominant in each $C_j^i$ in spite of $\Delta/H \ll 1$. 
}
\begin{center}
\begin{tabular}{l}
\hline 
Coefficient \\
\hline 
${\mbox{\boldmath $I$}}//[100]$ \\
$C_0^{[100]} = C_2^{[100]} - C_4^{[100]}$ \\
$C_2^{[100]}=\displaystyle{\frac{3}{8} \left( \frac{\lambda}{H} \right)^2
r_{s, + \to \varepsilon 2, - }}$ \\
$C_4^{[100]}=\displaystyle{\frac{3}{32} \left( \frac{\lambda}{H} \right)^2
(r_{s, + \to \varepsilon 2, - } -  r_{s, + \to \varepsilon 1, - })}$ 
\\\\[-0.3cm]
${\mbox{\boldmath $I$}}//[110]$ \\
$C_0^{[110]}=C_2^{[110]} - C_4^{[110]} + C_6^{[110]} - C_8^{[110]}$ \\
$C_2^{[110]}=\displaystyle{\frac{3}{8} \left( \frac{\lambda}{H} \right)^2
r_{s, + \to \varepsilon 1, - }}$ \\
$C_4^{[110]}=\displaystyle{\frac{3}{32} \left( \frac{\lambda}{H} \right)^2
(r_{s, + \to \varepsilon 1, - } -  r_{s, + \to \varepsilon 2, - })}$ \\
$C_6^{[110]}=0$ \\
$C_8^{[110]}=0$ 
\\\\[-0.3cm]
${\mbox{\boldmath $I$}}//[001]$ \\
$C_0^{[001]}=-C_4^{[001]}$ \\
$C_4^{[001]}=\displaystyle{\frac{3}{8}
\left( \frac{\lambda }{H} \right)^2 
(r_{s,+ \to \varepsilon 2, -} - r_{s,+ \to \varepsilon 1, -})}$  \\
\hline 
\end{tabular}
\end{center}
\label{table3}
\end{table}

\section{Consideration}
\label{sec_cons}

We consider 
the origin of $C_j^i \cos j \phi_i$ for type A 
and 
features of $C_j^i$ for types A, B, and C.

\subsection{Origin of $C_j^i \cos j \phi_i$ for type A} 
\label{origin}
We point out that 
$C_j^i \cos j\phi_i$ for type A 
originates from 
the changes 
in the d states 
[i.e., $|m,\chi_\varsigma (\phi))$ of Eq. (\ref{|m,chi_s)})] 
due to $V_{\rm so}$, 
where 
the changes 
are expressed by 
$c_{m,\varsigma} (\phi)$ and 
$c_{n,\sigma}^{m,\varsigma}(\phi)$ 
in 
Eq. (\ref{|m,chi_s)}). 
As shown in Eqs. (\ref{C_2^100_gamma})$-$(\ref{C_4^001_rho}), 
$C_j^i \cos j\phi_i$ 
has a single $\rho_{j,\sigma}^{i,(2)} \cos j\phi_i$ 
in the numerator of each term. 
This $\rho_{j,\sigma}^{i,(2)} \cos j \phi_i$ 
consists of the second-order terms of 
$\lambda/H$, $\lambda/\Delta$, 
$\lambda/(H \pm \Delta)$, 
$\delta_t/H$, $\delta_t/\Delta$, 
and $\delta_t/(H \pm \Delta)$, 
with $t=\varepsilon$ or $\gamma$ (see Appendix \ref{appen_resis}). 
The second-order terms 
are related to 
the changes in the d states due to $V_{\rm so}$ 
[see Eqs. (\ref{rho_sigma^i})$-$(\ref{tau_sd_inv}) and (\ref{|m,chi_s)})].

Table \ref{table4} shows 
the origin of 
$C_j^i \cos j \phi_i$ with $i=[100]$, $[110]$, and $[001]$ for type A. 
Here, we pay attention to the overlap integrals of 
Eqs. (\ref{s100_xy})$-$(\ref{s001_+}). 
We find that 
$C_2^i \cos 2\phi_i$ and $C_6^i \cos 6\phi_i$ 
are related to the probability amplitudes 
of the slightly hybridized states 
and 
$C_4^i \cos 4\phi_i$ is related to the probability of 
the slightly hybridized state (i.e., $|3z^2-r^2,\chi_\pm \rangle$). 
In addition, 
$C_8^i \cos 8\phi_i$ is related to the probability of 
the slightly hybridized state (i.e., $|xy,\chi_- \rangle$) 
and 
the probability amplitude 
of the slightly reduced state (i.e., $|xy,\chi_- \rangle$) 
in the dominant states. 
The details are explained in Appendix \ref{appen_origin}.

\begin{table*}[ht]
\caption{
Origin of 
$C_j^i \cos j\phi_i$ 
for type A in Table \ref{table1}, 
with $j$=2, 4, 6, and 8; 
$i=[100]$, $[110]$, and $[001]$; 
$\phi_{[100]}=\phi_{[001]}=\phi$; 
and $\phi_{[110]}=\phi'$. 
Type A has 
$D_{m,+}^{(d)}=0$, 
$D_{m,-}^{(d)}\ne 0$, 
$r_{s,\sigma \to \delta_\varepsilon,-} 
\equiv r_{s,\sigma \to \varepsilon 1, -}$, 
$r_{s,\sigma \to \xi_+,-}=r_{s,\sigma \to \xi_-,-} \equiv 
r_{s,\sigma \to \varepsilon 2, -}$, 
$r_{s,\sigma \to 3z^2-r^2, -} = r_{s,\sigma \to x^2-y^2, -} \equiv  
r_{s,\sigma \to \gamma, -}$, 
and $\Delta/H \ll 1$. 
The symbol PA (P) represents the probability amplitude (probability). 
The symbol N/A means ``not applicable''. 
For ``PA of $|xy,\chi_- \rangle$'' in $C_8^{[110]} \cos 8\phi_{[110]}$, 
this $|xy,\chi_- \rangle$ is the slightly reduced state, 
which is included in 
$|m,\chi_\varsigma (\phi) \rangle$ in Eq. (\ref{|m,chi_s)}). 
In contrast, 
the other states in this table 
are the slightly hybridized states, 
which are included in 
$|n,\chi_\sigma (\phi) \rangle$ in Eq. (\ref{|m,chi_s)}). 
}
\begin{center}
{\small 
\begin{tabular}{llllll}
\hline \\[-0.5cm]
$i$ & $C_2^i \cos 2\phi_i$ & $C_4^i \cos 4\phi_i$ & $C_6^i \cos 6 \phi_i$ & $C_8^i \cos 8\phi_i$ \\
\hline \\[-0.5cm]
 $[100]$ & PA of $|3z^2-r^2,\chi_\pm \rangle$ & P of $|3z^2-r^2,\chi_\pm \rangle$ & N/A & N/A \\
  & PA of $|x^2-y^2,\chi_- \rangle$ & 
&  &  \\
 $[110]$ & PA of $|3z^2-r^2,\chi_\pm \rangle$ & P of $|3z^2-r^2,\chi_\pm \rangle$ & PA of $|3z^2-r^2,\chi_- \rangle$ & P of $|xy,\chi_- \rangle$ \\
  & PA of $|xy,\chi_- \rangle$ &  &(PA of $|3z^2-r^2,\chi_- \rangle$)  & PA of $|xy,\chi_- \rangle$  \\
 & (PA of $|3z^2-r^2,\chi_- \rangle$) & & $\times$(PA of $|xy,\chi_- \rangle$)  &   \\
 &$\times$(PA of $|xy,\chi_- \rangle$) & &  &   \\
 $[001]$ & N/A & P of $|3z^2-r^2,\chi_\pm \rangle$ & N/A & N/A \\ 
\hline 
\end{tabular}
}
\end{center}
\label{table4}
\end{table*}

\subsection{Features in $C_j^i$ for types A, B, and C} 
We describe the features of the respective terms in $C_j^i$ 
for type A in Table \ref{table1}. 
We first find that 
$C_j^i$ consists of 
the terms with $(\lambda/\Delta)^2$, $(\lambda/H)^2$, 
and $\lambda^2/(H\Delta)$. 
Their terms are related to 
the changes in the d states 
due to $V_{\rm so}$, 
as noted in Sect. \ref{origin}. 
On the basis of 
$| (m,\chi_\varsigma (\phi)|
e^{i{\mbox{\boldmath $k$}}_\sigma^i 
\cdot {\mbox{\boldmath $r$}}},\chi_\sigma (\phi) \rangle |^2$ 
in Eq. (\ref{tau_sd_inv}), 
we show that such terms arise from 
the following two origins. 
One is 
the square of the first-order perturbation terms 
in the d states 
such as 
$\sum_{k (\ne m, m_i)} 
\frac{ V_{k,m}}{E_m -E_k} |k \rangle$
in Eq. (\ref{wf_general}), 
where the square comes from 
the above-mentioned square of the overlap integral. 
The other is the second-order perturbation terms 
in the d states 
such as 
$\sum_{k (\ne m, m_i)} \sum_{n (\ne m, m_i)}
\frac{ V_{n,m}}{E_m -E_n}
\frac{ V_{k,n}}{E_m -E_k} |k \rangle$ in Eq. (\ref{wf_general}). 
Specifically, 
the second-order perturbation terms are multiplied by 
the zero-order term [i.e., $|m \rangle$ in Eq. (\ref{wf_general})] 
in the calculation of 
the above-mentioned square of the overlap integral. 
Here, 
$H$ in the denominators 
is the energy difference between the different spin states. 
This $H$ therefore 
indicates the hybridization between them. 
In contrast, 
$\Delta$ in the denominators 
is the energy difference between the same spin states. 
This $\Delta$ 
represents the hybridization between them\cite{HD}. 
Next, using Eqs. (\ref{r_s-d}) and (\ref{proportional}), 
we confirm that 
$C_4^{[100]}$, $C_4^{[110]}$, $C_6^{[110]}$, $C_8^{[110]}$, 
and $C_4^{[001]}$ 
are proportional to 
$D_{\varepsilon 2,-}^{(d)} - D_{\varepsilon 1,-}^{(d)}$. 
Their magnitudes may indicate 
the degree of the tetragonal distortion. 
In addition, 
the signs of 
$C_6^{[110]}$ and $C_8^{[110]}$ 
reveal 
the magnitude relation of 
$D_{\varepsilon 2,-}^{(d)}$ and $D_{\varepsilon 1,-}^{(d)}$. 
Note also that 
all of the terms in the coefficients of type A 
are extracted for 
type B in Table \ref{table2} and type C in Table \ref{table3}.

For type B in Table \ref{table2}, 
we find that 
$C_4^{[100]}$, $C_4^{[110]}$, $C_6^{[110]}$, $C_8^{[110]}$, 
and $C_4^{[001]}$ 
are proportional to 
$D_{\varepsilon 2,-}^{(d)} - D_{\varepsilon 1,-}^{(d)}$. 
Their signs 
reveal 
the magnitude relation of 
$D_{\varepsilon 2,-}^{(d)}$ and $D_{\varepsilon 1,-}^{(d)}$.

For type C in Table \ref{table3}, 
we find that $C_2^{[100]}$ and $C_2^{[110]}$ 
are proportional to the PDOS 
of the $d\varepsilon$ states 
at $E_{\mbox{\tiny F}}$ 
[also see Eqs. (\ref{r_s-d}) and (\ref{proportional})]. 
Their signs are always positive. 
In contrast, 
$C_4^{[100]}$, $C_4^{[110]}$, and $C_4^{[001]}$ 
are proportional to 
$D_{\varepsilon 2,-}^{(d)} - D_{\varepsilon 1,-}^{(d)}$. 
Their signs 
indicate 
the magnitude relation of 
$D_{\varepsilon 2,-}^{(d)}$ and $D_{\varepsilon 1,-}^{(d)}$.

\section{
$C_4^{[100]} = -C_4^{[110]}$}
\label{C_4^100=-C_4^110}
We obtain the relation 
$C_4^{[100]} = -C_4^{[110]}$ of Eq. (\ref{relation}), 
which was experimentally observed for 
Ni\cite{Dedie,Ni_C4}, 
under the condition of 
$r_{s,- \to \varepsilon 2,-} = r_{s,- \to \gamma,-}$. 
The details are described below.

We first show the condition to obtain $C_4^{[100]} = -C_4^{[110]}$ 
on the basis of the features of 
$C_4^{[100]}$ and $C_4^{[110]}$. 
Under the condition of 
$r_{s,- \to x^2-y^2, -} = r_{s,- \to 3z^2-r^2, -} \equiv r_{s,- \to \gamma, -}$ of Eq. (\ref{gamma}) (i.e., $\rho_{s,- \to x^2-y^2,-} = \rho_{s,- \to 3z^2 -r^2, -}$ or $D_{x^2-y^2,-}^{(d)}=D_{3z^2-r^2,-}^{(d)}$), 
$C_4^{[100]}$ of Eq. (\ref{C_4^100_gamma}) 
consists of 
$\rho_{4,\pm}^{[100],(2)}$ of Eq. (3) in Ref. \citen{Kokado3_1} and 
$\rho_{0,\pm}^{[100],(0)}$ of Eq. (43) in Ref. \citen{Kokado3}, 
where 
$\rho_{s,\pm \to m,+}=0$ due to $D_{m,+}^{(d)}=0$, 
$\rho_{s,- \to x^2-y^2,-} = \rho_{s,- \to 3z^2 -r^2, -}$, 
and 
Eqs. (\ref{epsilon1}) and (\ref{epsilon2}) 
[i.e., Eqs. (\ref{rho_epsilon1}) and (\ref{rho_epsilon2})] are set. 
In addition, 
$C_4^{[110]}$ of Eq. (\ref{C_4^110_rho}) is composed of 
$\rho_{4,\pm}^{[110],(2)}$ of 
Eqs. (\ref{rho_4+^110(2)}) and (\ref{rho_4-^110(2)}) 
and $\rho_{0,\pm}^{[110],(0)}$ of 
Eqs. (\ref{rho_0+^110(0)}) and (\ref{rho_0-^110(0)}), 
where 
$\rho_{s,- \to x^2-y^2,-} = \rho_{s,- \to 3z^2 -r^2, -}$. 
In this case, 
$\rho_{4,\pm}^{[100],(2)}$ 
and $\rho_{4,\pm}^{[110],(2)}$ 
satisfy\cite{varepsilon12} 
\begin{eqnarray}
\label{rho_4_relation}
&&\rho_{4,\pm}^{[100],(2)}=-\rho_{4,\pm}^{[110],(2)}. 
\end{eqnarray}
Furthermore, under the condition of 
$r_{s,- \to \varepsilon 2,-} = r_{s,- \to x^2-y^2,-}$ 
(i.e., $D_{\xi +,-}^{(d)}=D_{\xi -,-}^{(d)}=
D_{x^2-y^2,-}^{(d)}$), 
$\rho_{0,\pm}^{[100],(0)}$ and $\rho_{0,\pm}^{[110],(0)}$ 
satisfy 
\begin{eqnarray}
\label{rho_0_relation}
&&\rho_{0,\pm}^{[100],(0)}=\rho_{0,\pm}^{[110],(0)}.
\end{eqnarray}
As a result, under the condition of 
$r_{s,- \to \varepsilon 2,-} = r_{s,- \to \gamma,-}$ 
(i.e., $D_{\xi +,-}^{(d)}=D_{\xi -,-}^{(d)}=
D_{x^2-y^2,-}^{(d)}=D_{3z^2-r^2,-}^{(d)}$)\cite{e-g}, 
we can obtain 
$C_4^{[100]} = -C_4^{[110]}$ 
using Eqs. (\ref{rho_0_relation}), (\ref{rho_4_relation}), 
(\ref{C_4^100_gamma}), and (\ref{C_4^110_rho}). 
Here, 
Eq. (\ref{rho_0_relation}) represents 
the equality between the constant terms, which are independent of $\phi$. 
In contrast, 
Eq. (\ref{rho_4_relation}) directly contributes to 
$C_4^{[100]} = -C_4^{[110]}$. 

We next explain the relation of Eq. (\ref{rho_4_relation}) in detail. 
For $\rho_{4,\pm}^{[100],(2)}$ and $\rho_{4,\pm}^{[110],(2)}$, 
we consider 
$\rho_{4,\pm}^{[100],(2)} \cos4\phi$ 
in Eqs. (\ref{rho_pm^100}) and (\ref{rho_4pm^100}) 
and $\rho_{4,\pm}^{[110],(2)} \cos 4\phi'$ 
in Eqs. (\ref{rho_pm^110}) and (\ref{rho_4pm^110}). 
They originally arise from the overlap integrals 
between the plane wave and $|3z^2-r^2,\chi_\pm(\phi) \rangle$ 
(also see Table \ref{table4}), 
where this $|3z^2-r^2,\chi_\pm(\phi) \rangle$ is included 
in 
$|\xi_+,\chi_- (\phi) )$ 
and 
$|\delta_\varepsilon,\chi_- (\phi))$. 
The overlap integral for ${\mbox{\boldmath $I$}}//[100]$ 
is given by Eq. (\ref{s100_3z2-r2}), 
and that for ${\mbox{\boldmath $I$}}//[110]$ 
is given by Eq. (\ref{s110_3z2-r2}). 
We emphasize here that 
Eqs. (\ref{s100_3z2-r2}) and (\ref{s110_3z2-r2}) 
produce the same expression 
in spite of the difference in the plane waves 
between ${\mbox{\boldmath $I$}}//[100]$ and ${\mbox{\boldmath $I$}}//[110]$. 
This feature reflects the fact 
that $|3z^2-r^2,\chi_\pm(\phi) \rangle$ 
possesses continuous rotational symmetry 
around the $z$-axis. 
As a result, 
the ${\mbox{\boldmath $I$}}//[100]$ and ${\mbox{\boldmath $I$}}//[110]$ cases 
give 
the same fourfold symmetric resistivity, 
$\rho_{\pm}^* \cos4 \phi$, 
where $\rho_+^*$ ($\rho_-^*$) represents 
the coefficient of the up (down) spin 
of the $\cos 4\phi$ term. 
When ${\mbox{\boldmath $I$}}//[100]$, 
we have 
$\rho_{\pm}^* \cos 4\phi$$\equiv$$\rho_{4,\pm}^{[100],(2)} \cos4\phi$, 
i.e., $\rho_{\pm}^* \equiv \rho_{4,\pm}^{[100],(2)}$. 
When ${\mbox{\boldmath $I$}}//[110]$, 
we obtain $\rho_\pm^* \cos 4\phi = -\rho_\pm^* \cos 4\phi'$ 
by substituting $\phi=\phi' + \pi/4$ into $\rho_\pm^* \cos 4\phi$. 
We then have 
$- \rho_{\pm}^* \cos 4\phi' 
\equiv \rho_{4,\pm}^{[110],(2)} \cos 4\phi'$, 
i.e., $-\rho_{\pm}^* \equiv \rho_{4,\pm}^{[110],(2)}$. 
The above results thus 
give 
the relation of 
$\rho_{4,\pm}^{[100],(2)} =-\rho_{4,\pm}^{[110],(2)}$ 
of Eq. (\ref{rho_4_relation}).

We also mention that Eq. (\ref{relation}) is found 
in the expression 
by D$\ddot{\rm o}$ring,\cite{Doring} 
i.e., Eqs. (\ref{C4_100_D}) and (\ref{C4_110_D}). 
It is noted here that 
D$\ddot{\rm o}$ring's expression  
[i.e., Eq. (\ref{Doring})] does not directly need 
the condition of 
$r_{s,- \to x^2-y^2, -} = r_{s,- \to 3z^2-r^2, -}$ 
and 
$r_{s,- \to \varepsilon 2,-} = r_{s,- \to x^2-y^2,-}$ 
to obtain Eq. (\ref{relation}). 
In other words, 
such a condition appears to be originally included 
in D$\ddot{\rm o}$ring's expression. 
First, 
$\Delta \rho/\rho$ of 
Eq. (\ref{Doring}) is an expression for the cubic system 
and this system exhibits 
$r_{s,- \to x^2-y^2, -} = r_{s,- \to 3z^2-r^2, -}$ 
due to $D_{x^2-y^2,-}^{(d)}=D_{3z^2-r^2,-}^{(d)}$. 
Next, 
the condition of $r_{s,- \to \varepsilon 2,-} = r_{s,- \to x^2-y^2,-}$ 
comes from 
a constant term, $C_0$, 
in the expression for the resistivity in Ref. \citen{McGuire1}, 
where $C_0$ 
is independent of the current direction. 
The constant term $C_0$ corresponds only to 
$\rho_{0,+}^{[100],(0)} \rho_{0,-}^{[100],(0)} 
/(\rho_{0,+}^{[100],(0)} + \rho_{0,-}^{[100],(0)})$ 
or 
$\rho_{0,+}^{[110],(0)} \rho_{0,-}^{[110],(0)} 
/(\rho_{0,+}^{[110],(0)} + \rho_{0,-}^{[110],(0)})$ 
in the present theory, 
where 
$\rho_{s,\pm \to m,+}=0$ due to $D_{m,+}^{(d)}=0$ 
should be set for 
$\rho_{0,\pm}^{[100],(0)}$ of 
Eq. (43) in Ref. \citen{Kokado3}. 
Here, $\rho_{0,\pm}^{[100],(0)}$ and $\rho_{0,\pm}^{[110],(0)}$ 
consist of 
$\rho_{s,\pm \to \varepsilon 2,-}$ and $\rho_{s,\pm \to x^2-y^2,-}$, 
respectively. 
The other parts 
in $\rho_{0,\pm}^{[100],(0)}$ and $\rho_{0,\pm}^{[110],(0)}$ 
are equal. 
From 
$C_0 = \rho_{0,+}^{[100],(0)} \rho_{0,-}^{[100],(0)} 
/(\rho_{0,+}^{[100],(0)} + \rho_{0,-}^{[100],(0)}) = 
\rho_{0,+}^{[110],(0)} \rho_{0,-}^{[110],(0)} 
/(\rho_{0,+}^{[110],(0)} + \rho_{0,-}^{[110],(0)})$, 
we therefore obtain 
$\rho_{s,- \to \varepsilon 2,-} = \rho_{s,- \to x^2-y^2,-}$, 
i.e., $r_{s,- \to \varepsilon 2,-} = r_{s,- \to x^2-y^2,-}$ 
[see Eq. (\ref{r_s-d})].

\section{Coefficients for Ni}
\label{sec_Ni}
Using $C_j^i$ for type A, 
we qualitatively explain 
the experimental results of 
$C_2^{[100]}$, $C_4^{[100]}$, $C_2^{[110]}$, and $C_4^{[110]}$ 
at 293 K for Ni (see Table \ref{table5}). 
In particular, 
we focus on their signs. 
The details are described below. 


We first note that 
the experimental values in Table \ref{table5} 
indicate 
the estimated values of $C_j^i$ 
in the expression for the AMR ratio by D$\ddot{\rm o}$ring 
in Appendix \ref{appen_pheno}. 
These values are estimated 
by applying 
D$\ddot{\rm o}$ring's expression to 
the experimentally observed AMR ratio.\cite{Dedie} 
Here, 
D$\ddot{\rm o}$ring's expression 
consists of $C_j^i \cos j\phi_i$ with $j=$0, 2, and 4; 
that is, 
this expression 
does not take into account 
higher-order terms 
of $C_j^i \cos j \phi_i$ with $j \ge 6$. 
In contrast, 
our theory 
produces 
higher-order terms 
of $C_j^{[110]} \cos j \phi'$ with $j \ge 6$. 
They are straightforwardly obtained 
up to the second order of 
$\lambda/H$, $\lambda/\Delta$, 
$\lambda/(H \pm \Delta)$, 
$\delta_t/H$, $\delta_t/\Delta$, 
and $\delta_t/(H \pm \Delta)$, 
with $t=\varepsilon$ or $\gamma$, 
where 
$(\frac{\lambda}{H - \Delta})^2 \approx \left( \frac{\lambda}{H} \right)^2$ 
and 
$\frac{\lambda^2}{\Delta (H - \Delta)} \approx \frac{\lambda^2}{H\Delta}
+ \left( \frac{\lambda}{H} \right)^2$ are also used. 



Next, 
as a model to investigate the experimental result, 
we choose 
type A in Table \ref{table1}. 
The procedure for choosing type A is as follows:

\begin{enumerate}
\item[(i)] 
The relatively large value of 
$C_4^{[100]}$ ($=-C_4^{[110]}$) in Table \ref{table5} 
means that this Ni has 
the crystal field of tetragonal symmetry, 
as found from the calculation results 
using the exact diagonalization method 
(see Figs. 7 and 8 in Ref. \citen{Kokado3}).\cite{others} 
We therefore adopt the present model 
with the crystal field of tetragonal symmetry.

\item[(ii)] 
We set the condition of 
$r_{s,- \to \varepsilon 2, -}=r_{s,- \to \gamma, -}$ 
to reproduce $C_4^{[100]} = -C_4^{[110]}$ 
(see Sect. \ref{C_4^100=-C_4^110}). 
We note here that 
this condition 
may be 
valid for Ni. 
First, we believe that 
$r_{s,- \to x^2-y^2, -} = r_{s,- \to 3z^2-r^2, -} \equiv  
r_{s,- \to \gamma, -}$ of Eq. (\ref{gamma}) 
(i.e., $D_{x^2-y^2,-}^{(d)}=D_{3z^2-r^2,-}^{(d)}$) 
has no problem 
when 
$\delta_\gamma$ is relatively small as shown in Fig. \ref{sample}. 
We next consider the condition of 
$r_{s,- \to \varepsilon 2,-} = r_{s,- \to \gamma,-}$ 
(i.e., $\rho_{s,- \to \varepsilon 2,-} = \rho_{s,- \to \gamma,-}$). 
From Ref. \citen{e-g}, 
we know that 
the condition 
may be rewritten as 
$\left| \frac{(3/4)(\rho_{s,- \to \varepsilon 2,-} - \rho_{s,- \to \gamma,-})}{\rho_{s,-}+ \rho_{s,- \to \gamma,-}} \right| \ll 1$. 
This is now expressed as 
$\left| \frac{(3/4)(r_{s,- \to \varepsilon 2,-} - r_{s,- \to \gamma,-})}{r+ r_{s,- \to \gamma,-}} \right| \ll 1$ 
by using Eqs. (\ref{r_s-s}) and (\ref{r_s-d}). 
As noted below, 
we choose 
$r=3.00$ and $r_{s,- \to \gamma,-}=2.00$ for Ni. 
In this case, 
we have 
$\left| r_{s,- \to \varepsilon 2,-} - r_{s,- \to \gamma,-} \right| \ll 6.67$. 
On the other hand, 
$r_{s,- \to \varepsilon 2,-}$ and $r_{s,- \to \gamma,-}$ 
were evaluated to be about 2.5 for Ni in a previous study\cite{Kokado1}. 
We therefore roughly estimate 
$\left| r_{s,- \to \varepsilon 2,-} - r_{s,- \to \gamma,-} \right| < 1$. 
This inequality satisfies 
$\left| r_{s,- \to \varepsilon 2,-} - r_{s,- \to \gamma,-} \right| \ll 6.67$.

\item[(iii)]
We choose a type suitable for explaining the experimental result 
from types A, B, and C. 
Since the dominant $s$--$d$ scattering for Ni
is considered to be 
$s+ \to d-$\cite{Kokado2}, 
type C is a prime candidate, 
while type B is not a candidate. 
In type C in Table \ref{table3}, however, 
we cannot explain the experimental results 
in Table \ref{table5}. 
The reason is that 
the relation of 
$r_{s,+ \to \epsilon 2,-} > r_{s,+ \to \epsilon 1,-}$ 
deduced from 
the experimental result of 
$C_4^{[100]}>0$ and $C_4^{[110]}<0$ 
contradicts 
that of 
$r_{s,+ \to \epsilon 2,-} < r_{s,+ \to \epsilon 1,-}$ 
deduced from 
the experimental result of 
$C_2^{[110]}>C_2^{[100]}>0$. 
We thus choose type A, which is the comprehensive type. 
\end{enumerate}

For $C_j^i$ of type A, we roughly determine the parameters. 
From the previously evaluated 
$\rho_{s \to d\downarrow}/\rho_{s\uparrow}$ $(\sim 2.5)$ \cite{Kokado1}, 
we first set 
$r_{s,\pm \to \varepsilon 1, - } = 2.50$ 
and $r_{s,\pm \to \varepsilon 2, - } = r_{s,\pm \to \gamma,-}=2.00$, 
where 
the relation of 
$r_{s,\pm \to \varepsilon 1, - } > r_{s,\pm \to \varepsilon 2, - } (= r_{s,\pm \to \gamma,-})$ results in $C_2^{[110]}>C_2^{[100]}$. 
We next choose the other parameters 
so as to reproduce the experimental results, to some extent: 
$\lambda/H= 1.10 \times 10^{-1}$,\cite{H_Ni} 
$H/\Delta=7.00$, 
and $r=3.00$\cite{rrr}. 

Table \ref{table5} shows the theoretical values of $C_j^i$. 
The theoretical values of $C_2^i$ and $C_4^i$ agree qualitatively with 
the respective experimental ones. 
Namely, 
the signs of the theoretical values of $C_2^i$ and $C_4^i$ 
are the same as the respective experimental ones. 
The theoretical value of $C_2^i$ is relatively close to 
its experimental one. 
On the other hand, 
the theoretical value of $|C_4^i|$ 
is considerably different from its experimental one. 
In addition, our theory gives $C_6^{[110]}$ and $C_8^{[110]}$, 
which were not evaluated in the experiment. 
The relation of 
$C_2^{[110]} > |C_6^{[110]}| > |C_8^{[110]}| > |C_4^{[110]}|$ 
is also obtained in our theory. 
Such a difference between the experimental and theoretical results 
may be a future subject of research. 
In particular, 
$C_6^{[110]}$ and $C_8^{[110]}$ 
may be evaluated 
by extending D$\ddot{\rm o}$ring's expression 
to the expression with higher-order terms 
of $C_j^{[110]} \cos j \phi'$ with $j \ge 6$\cite{expansion} 
and applying the extended expression to the experimental result. 
Our theoretical values of $C_6^{[110]}$ and $C_8^{[110]}$ 
may be then examined on the basis of the experimentally evaluated values. 


We discuss the dominant $s$--$d$ scatterings 
observed in 
$C_2^{[100]}$ ($>0$), $C_4^{[100]}$ ($>0$), 
$C_2^{[110]}$ ($>0$), and $C_4^{[110]}$ ($<0$) in Table \ref{table1}. 
Here, we focus on the dominant terms in 
$C_2^{[100]}$, $C_4^{[100]}$, 
$C_2^{[110]}$, and $C_4^{[110]}$. 
The dominant terms in $C_2^{[100]}$ and $C_2^{[110]}$ 
are, respectively, the terms with 
$\left( \frac{\lambda}{H} \right)^2 
r_{s, + \to \varepsilon 2, - }(r + r_{s,- \to \gamma,-})$ 
and 
$\left( \frac{\lambda}{H} \right)^2 
r_{s, + \to \varepsilon 1, - }
\left(r + \frac{3}{4}r_{s,- \to \varepsilon 2,-}+ \frac{1}{4}r_{s,- \to \gamma,-} \right)$, 
which are positive. 
These terms arise from the $s$--$d$ scattering ``$s,+ \to d,-$''. 
In contrast, 
the dominant term 
in $C_4^{[100]}$ 
is the term with $(\lambda/\Delta)^2$, which is positive. 
In addition, the dominant term in $C_4^{[110]}$ 
is the term with $(\lambda/\Delta)^2$, which is negative. 
These terms arise from the $s$--$d$ scattering ``$s,- \to d,-$''. 
The dominant $s$--$d$ scattering observed in $C_4^i$ 
is thus different from that observed in $C_2^i$.

We also comment on 
the dominant terms in $C_4^{[100]}$ and $C_4^{[110]}$, 
i.e., the terms with $(\lambda/\Delta)^2$.  
When 
$C_4^{[100]}$ and $C_4^{[110]}$ 
are approximated as 
only the terms with $(\lambda/\Delta)^2$, 
we have 
\begin{eqnarray}
\label{C_4_DOS_relation}
&&C_4^{[100]}= - C_4^{[110]} \propto 
\left( \frac{\lambda}{\Delta} \right)^2
\left( D_{\varepsilon 1,-}^{(d)}  - D_{\varepsilon 2,-}^{(d)} \right), 
\end{eqnarray}
where 
Eqs. (\ref{r_s-d}) and (\ref{proportional}) are used. 
From Eq. (\ref{C_4_DOS_relation}) and the experimental results 
of $C_4^{[100]}>0$ and $C_4^{[110]}<0$, 
we predict 
the relation of $D_{\varepsilon 1,-}^{(d)} > D_{\varepsilon 2,-}^{(d)}$ 
due to the tetragonal distortion.

\begin{table}[ht]
\caption{
The experimental values of 
$C_2^{[100]}$, 
$C_4^{[100]}$, 
$C_2^{[110]}$, and $C_4^{[110]}$ 
at 293 K for Ni\cite{Dedie,Ni_C4} 
and the theoretical values calculated from 
$C_j^i$ for type A in Table \ref{table1}. 
In this calculation, we use 
the parameter sets of 
$r_{s,\pm \to \varepsilon 1, - } = 2.50$\cite{Kokado1}, 
$r_{s,\pm \to \varepsilon 2, - } =
r_{s,\pm \to \gamma, - } =2.00$\cite{Kokado1}, 
$\lambda/H= 1.10 \times 10^{-1}$\cite{H_Ni}, 
$H/\Delta=7.00$, 
and $r=3.00$\cite{rrr}. 
}
\begin{center}
{\small 
\begin{tabular}{lcccccc}
\hline 
& $C_2^{[100]}$ & $C_4^{[100]}$ & $C_2^{[110]}$ & $C_4^{[110]}$ & 
$C_6^{[110]}$
& 
$C_8^{[110]}$
\\
\hline 
Experiment\cite{Dedie,Ni_C4} & $5.00 \times 10^{-3}$ & $2.63 \times 10^{-3}$ & $1.25 \times 10^{-2}$ & $-2.63 \times 10^{-3}$ & 
-
& 
-
\\
Theory & $3.55\times 10^{-3}$ & $4.54 \times 10^{-4}$ & $1.24 \times 10^{-2}$ &  $-4.54 \times 10^{-4}$ & 
$-3.23 \times 10^{-3}$
& 
$-2.81 \times 10^{-3}$
\\
\hline 
\end{tabular}
}
\end{center}
\label{table5}
\end{table}

\section{Conclusion}
\label{sec_conc}
We theoretically studied 
AMR$^{[100]}(\phi)$, 
AMR$^{[110]}(\phi')$, 
and 
AMR$^{[001]}(\phi)$ 
for ferromagnets 
with the crystal field of tetragonal symmetry. 
Here, we used the electron scattering theory 
for a system consisting of the conduction electron state and the localized d states. 
The d states were obtained by using the perturbation theory. The main results are as follows: 
\begin{enumerate}
\item[(i)] 
We derived expressions for 
AMR$^{[100]}(\phi)$, 
AMR$^{[110]}(\phi')$, 
and 
AMR$^{[001]}(\phi)$ for ferromagnets 
with $D_{m,+}^{(d)}=0$ and $D_{m,-}^{(d)} \ne 0$. 
The coefficient 
$C_j^i$ 
is composed of 
$\lambda$, $H$, $\Delta$, $\delta_\varepsilon$, $\delta_\gamma$, 
and $s$--$s$ and $s$--$d$ resistivities. 
From such $C_j^i$, 
we obtained a simple expression for $C_j^i$ 
for the simplified system 
with 
$r_{s,- \to 3z^2-r^2, -} = r_{s,- \to x^2-y^2, -}$. 
This system was divided into types A, B, and C. 
Type A is the generalized strong ferromagnet 
with the $s$--$d$ scattering ``$s,+ \to d,-$ and $s,- \to d,-$'', 
type B is the half-metallic ferromagnet 
with the dominant $s$--$d$ scattering ``$s,- \to d,-$'', 
and 
type C is the specified strong ferromagnet 
with the $s$--$d$ scattering ``$s,+ \to d,-$''. 
The coefficient $C_j^i$ for type A 
includes $C_j^i$ for types B and C. 
The AMR ratios 
AMR$^{[100]}(0)$ and AMR$^{[110]}(0)$ 
for type A 
also corresponded to 
that of 
the CFJ model\cite{Campbell} 
under the condition of the CFJ model.

\item[(ii)] 
We found that 
$C_j^i \cos j\phi_i$ for type A 
originates from 
the changes 
in the d states due to $V_{\rm so}$. 
Concretely, 
$C_j^i \cos j\phi_i$ is related to 
the probability amplitudes 
and probabilities of the slightly hybridized states 
or 
the probability amplitudes 
of the slightly reduced state in the dominant states.

\item[(iii)] 
For type A, 
$C_j^i$ has terms with $(\lambda/\Delta)^2$, $(\lambda/H)^2$, 
and $\lambda^2/(H\Delta)^2$. 
In addition, 
$C_4^{[100]}$, $C_4^{[110]}$, $C_6^{[110]}$, $C_8^{[110]}$, 
and $C_4^{[001]}$ 
are proportional to 
$D_{\varepsilon 2,-}^{(d)} - D_{\varepsilon 1,-}^{(d)}$. 
Their magnitudes may indicate 
the degree of the tetragonal distortion. 
For type B, 
$C_4^{[100]}$, $C_4^{[110]}$, $C_6^{[110]}$, $C_8^{[110]}$, 
and $C_4^{[001]}$ 
are proportional to 
$D_{\varepsilon 2,-}^{(d)} - D_{\varepsilon 1,-}^{(d)}$. 
Their signs 
reveal 
the magnitude relation of 
$D_{\varepsilon 2,-}^{(d)}$ and $D_{\varepsilon 1,-}^{(d)}$.
For type C, 
$C_j^i$ is proportional to $(\lambda/H)^2$. 
In addition, 
$C_2^{[100]}$ and $C_2^{[110]}$ 
are proportional to the PDOS 
of the $d\varepsilon$ states at $E_{\mbox{\tiny F}}$. 
In contrast, 
$C_4^{[100]}$, $C_4^{[110]}$, and $C_4^{[001]}$ 
are proportional to 
$D_{\varepsilon 2,-}^{(d)} - D_{\varepsilon 1,-}^{(d)}$. 
Their signs 
indicate 
the magnitude relation of 
$D_{\varepsilon 2,-}^{(d)}$ and $D_{\varepsilon 1,-}^{(d)}$.

\item[(iv)] 
We obtained the relation $C_4^{[100]} = -C_4^{[110]}$ 
of Eq. (\ref{relation}) 
under the condition of 
$D_{\xi_+,-}^{(d)}=D_{\xi_-,-}^{(d)}=D_{x^2-y^2,-}^{(d)}=D_{3z^2-r^2,-}^{(d)}$. 
This relation could be explained by considering that 
$C_4^{[100]}$ and $C_4^{[110]}$ 
arise from the overlap integrals 
between the plane wave and $|3z^2-r^2,\chi_\pm(\phi) \rangle$, 
and 
the overlap integrals 
produce the same expression 
in spite of the difference in the plane waves 
between ${\mbox{\boldmath $I$}}//[100]$ and ${\mbox{\boldmath $I$}}//[110]$.

\item[(v)]
Using the expressions for $C_j^i$ for type A, 
we 
qualitatively explained 
the experimental results of 
$C_2^{[100]}$, $C_4^{[100]}$, $C_2^{[110]}$, and $C_4^{[110]}$ 
at 293 K for Ni. 
We found that 
the dominant $s$--$d$ scattering observed in $C_2^{[100]}$ and $C_2^{[110]}$ 
is $s,+ \to d,-$, 
while 
that observed in $C_4^{[100]}$ and $C_4^{[110]}$ 
is $s,- \to d,-$. 
From the experimental results 
of $C_4^{[100]}>0$ and $C_4^{[110]}<0$, 
we also predicted 
the relation of $D_{\varepsilon 1,-}^{(d)} > D_{\varepsilon 2,-}^{(d)}$ 
due to the tetragonal distortion. 
\end{enumerate}

\acknowledgments
This work has been supported by 
the Cooperative Research Project (H26/A04) of 
the RIEC, Tohoku University, 
and 
a Grant-in-Aid for Scientific Research (C) (No. 25390055) 
from the Japan Society for the Promotion of Science.

\appendix

\section{Expression for AMR Ratio by D$\ddot{\rm o}$ring}
\label{appen_pheno}
We report the expression for the AMR ratio by D$\ddot{\rm o}$ring, 
which consists of 
the expression for the resistivity 
based on the symmetry of a crystal.\cite{Doring,Bozorth,Gorkom} 
Here, we note that 
this expression is the same form as 
an expression for a spontaneous magnetostriction, 
which minimizes the total energy 
consisting of the magnetoelastic energy for a spin pair model 
and the elastic energy for a cubic system.\cite{Chikazumi}

The AMR ratio $\Delta \rho/\rho$ 
is expressed as 
\begin{eqnarray}
\label{Doring}
&&\frac{\Delta \rho}{\rho}= \frac{\rho - \rho_0}{\rho_0}\nonumber \\
&&\hspace*{0.8cm}= k_1 
\left( \alpha_1^2 \beta_1^2 + \alpha_2^2 \beta_2^2 + \alpha_3^2 \beta_3^2 - \frac{1}{3} \right) 
+2 k_2 
(\alpha_1 \alpha_2 \beta_1 \beta_2 + \alpha_2 \alpha_3 \beta_2 \beta_3
+\alpha_3 \alpha_1 \beta_3 \beta_1) \nonumber \\
&&\hspace*{1.2cm}+k_3
\left( \alpha_1^2 \alpha_2^2 + \alpha_2^2 \alpha_3^2 + \alpha_3^2 \alpha_1^2 
-\frac{1}{3} \right) \nonumber \\
&&\hspace*{1.2cm}+ k_4
\left[ 
\alpha_1^4 \beta_1^2 + \alpha_2^4 \beta_2^2 + \alpha_3^4 \beta_3^2
+ \frac{2}{3} 
\left( \alpha_1^2 \alpha_2^2 + \alpha_2^2 \alpha_3^2 + \alpha_3^2 \alpha_1^2 \right) - \frac{1}{3} \right] \nonumber \\
&&\hspace*{1.2cm}+ 2 k_5 
(\alpha_1 \alpha_2 \beta_1 \beta_2 \alpha_3^2 
+\alpha_2 \alpha_3 \beta_2 \beta_3 \alpha_1^2 
+\alpha_3 \alpha_1 \beta_3 \beta_1 \alpha_2^2 ), 
\end{eqnarray}
where 
$\rho$ is the resistivity for 
certain directions of 
${\mbox{\boldmath $I$}}$ and ${\mbox{\boldmath $M$}}$; 
$\rho_0$ is the average resistivity for the demagnetized state; 
$\alpha_1$, $\alpha_2$, and $\alpha_3$ indicate 
the direction cosines of 
the ${\mbox{\boldmath $M$}}$ direction; 
$\beta_1$, $\beta_2$, and $\beta_3$ denote 
the direction cosines of 
the ${\mbox{\boldmath $I$}}$ direction; 
and 
$k_1$, $k_2$, $k_3$, $k_4$, and $k_5$ are 
the coefficients.\cite{Doring,Bozorth} 
In this study, 
${\mbox{\boldmath $M$}}$ lies in the (001) plane (see Fig. \ref{sample}); 
that is, $\alpha_3$ is set to be 0.

The AMR ratio of ${\mbox{\boldmath $I$}}//[100]$, 
$(\Delta \rho/\rho)_{[100]}$, 
is obtained 
by substituting 
$(\alpha_1, \alpha_2, \alpha_3)=(\cos \phi, \sin \phi, 0)$ 
and $(\beta_1, \beta_2, \beta_3)=(1, 0, 0)$
into Eq. (\ref{Doring}):
\begin{eqnarray}
\label{Doring_100}
\left(\frac{\Delta \rho}{\rho}\right)_{[100]}
= C_0^{[100]} + C_2^{[100]} \cos 2\phi + C_4^{[100]} \cos 4\phi, 
\end{eqnarray}
with 
\begin{eqnarray}
&&C_0^{[100]}= \frac{1}{6}k_1 - \frac{5}{24}k_3 + \frac{1}{8}k_4, \\
\label{C2_100_D}
&&C_2^{[100]}= \frac{1}{2}k_1 + \frac{1}{2}k_4, \\
\label{C4_100_D}
&&C_4^{[100]}= -\frac{1}{8}k_3 + \frac{1}{24}k_4. 
\end{eqnarray}

The AMR ratio of ${\mbox{\boldmath $I$}}//[110]$, 
$(\Delta \rho/\rho)_{[110]}$, 
is obtained 
by substituting 
$(\alpha_1, \alpha_2, \alpha_3)=(\cos (\phi'+\pi/4), \sin (\phi'+\pi/4), 0)$ 
and 
$(\beta_1, \beta_2, \beta_3)=(1/\sqrt{2}, 1/\sqrt{2}, 0)$ 
into 
Eq. (\ref{Doring}):
\begin{eqnarray}
\label{Doring_110}
\left(\frac{\Delta \rho}{\rho}\right)_{[110]}
= C_0^{[110]} + C_2^{[110]} \cos 2\phi' + C_4^{[110]} \cos 4\phi', 
\end{eqnarray}
with 
\begin{eqnarray}
&&C_0^{[110]}=\frac{1}{6}k_1 - \frac{5}{24}k_3 + \frac{1}{8}k_4, \\
\label{C2_110_D}
&&C_2^{[110]}=\frac{1}{2}k_2, \\
\label{C4_110_D}
&&C_4^{[110]}=\frac{1}{8}k_3 - \frac{1}{24}k_4. 
\end{eqnarray}

The AMR ratio of ${\mbox{\boldmath $I$}}//[001]$, 
$(\Delta \rho/\rho)_{[001]}$, 
is obtained 
by substituting 
$(\alpha_1, \alpha_2, \alpha_3)=(\cos \phi, \sin \phi, 0)$ 
and $(\beta_1, \beta_2, \beta_3)=(0, 0, 1)$ 
into 
Eq. (\ref{Doring}):
\begin{eqnarray}
\label{Doring_001}
\left(\frac{\Delta \rho}{\rho}\right)_{[001]}
= C_0^{[001]} + C_4^{[001]} \cos 4\phi, 
\end{eqnarray}
with 
\begin{eqnarray}
&&C_0^{[001]}=
-\frac{1}{3}k_1 - \frac{5}{24}k_3 - \frac{1}{4}k_4, \\ 
&&C_4^{[001]}=
-\frac{1}{8}k_3 - \frac{1}{12}k_4. 
\end{eqnarray}

From Eqs. (\ref{C4_100_D}) and (\ref{C4_110_D}), 
we confirm the relation $C_4^{[100]} = -C_4^{[110]}$. 
In addition, 
$(\Delta \rho/\rho)_{[001]}$ of Eq. (\ref{Doring_001}) 
does not include the $\cos 2\phi$ term.

\section{Wave Function by Perturbation Theory 
for a Model with Degenerate Unperturbed Systems} 
\label{appen_wf}
We give an expression for the wave function 
$|m )$ 
of the first- and second-order perturbation theory 
for the case that 
the unperturbed system is degenerate. 
Here, $|m )$ is an abbreviated form
for $|m,\chi_\varsigma (\phi) )$ in Eq. (\ref{|m,chi_s)}). 
In this study, 
using this $|m )$, 
we 
obtain the wave functions 
from the matrix of ${\cal H}$ in Table I in Ref. \citen{Kokado3} 
(also see 
Sects. \ref{sub_Ham} and \ref{sec_d}).

We first give an eigenvalue equation 
for the unpertubed Hamiltonian ${\cal H}_0$ as
\begin{eqnarray}
\label{eigen_eq}
&&{\cal H}_0 |m_i \rangle = E_m |m_i \rangle, 
\end{eqnarray}
for $i=1$ - $N_m$. 
Here, $E_m$ is the eigenvalue 
and $|m_i \rangle$ ($i=1$ - $N_m$) is the eigenstate 
with the $N_m$-fold degeneracy.

On the basis of Eq. (\ref{eigen_eq}), 
we next derive the expression for $|m )$. 
When a specific state in $|m_i \rangle$ is written as 
$|m \rangle$, 
$|m )$ becomes

\newpage
\begin{eqnarray}
\label{wf_general}
&&|m ) = 
\left\{ 
1 - \left[ 
\frac{1}{2}
\displaystyle{\sum_{k (\ne m_i)}} 
\left| 
\frac{ V_{k,m}}{E_m -E_k}
\right|^2 
+ \frac{1}{2}
\displaystyle{\sum_{m_i' (\ne m)}} 
\left| 
\displaystyle{\sum_{n (\ne m_i)}} 
\frac{ V_{n,m}}{E_m -E_n} 
\frac{ V_{m_i',n}}{V_{m,m} -V_{m_i',m_i'}} 
\right|^2 
\right] \right\} |m \rangle \nonumber \\
&&\hspace*{1.2cm}+ \displaystyle{\sum_{k (\ne m_i)}} 
\frac{ V_{k,m}}{E_m -E_k} |k \rangle
+ \displaystyle{\sum_{m_i' (\ne m)}}
\left(
\displaystyle{\sum_{n (\ne m_i)}}
\frac{ V_{n,m}}{E_m -E_n} 
\frac{ V_{m_i',n}}{V_{m,m} -V_{m_i',m_i'}} 
\right) |m_i' \rangle \nonumber \\
&& \hspace*{1.2cm}
+ \displaystyle{\sum_{k (\ne m_i)}} 
\left[
\displaystyle{\sum_{n (\ne m_i)}} 
\frac{ V_{n,m}}{E_m -E_n}
\frac{ V_{k,n}}{E_m -E_k}
- 
\frac{ V_{m,m}V_{k,m}}{(E_m -E_k)^2} \right. \nonumber \\
&& \hspace*{1.2cm}\left. 
+ \displaystyle{\sum_{m_i' (\ne m)}} 
\left( 
\displaystyle{\sum_{n (\ne m_i)}} 
\frac{ V_{n,m}}{E_m -E_n}
\frac{ V_{m_i',n}}{V_{m,m} -V_{m_i',m_i'}} 
\right)
\frac{ V_{k,m_i'}}{E_m -E_k}
\right] |k \rangle \nonumber \\
&&\hspace*{1.2cm}
+ \displaystyle{\sum_{m_i'' (\ne m)}} 
\frac{ 1}{V_{m,m} -V_{m_i'',m_i''}} 
\left\{
\displaystyle{\sum_{n_1 (\ne m_i)}} 
V_{m_i'',n_1}
\left[
\displaystyle{\sum_{\ell_1 (\ne m_i)}} 
\frac{ V_{\ell_1,m}}{E_m -E_{\ell_1}}
\frac{ V_{n_1,\ell_1}}{E_m -E_{n_1}}
\right.\right. \nonumber \\
&&\hspace*{1.2cm} \left.\left.
- 
\frac{ V_{m,m}V_{n_1,m}}{(E_m -E_{n_1})^2} 
+ 
\displaystyle{\sum_{m_i' (\ne m)}} 
\left( 
\displaystyle{\sum_{\ell_2 (\ne m_i)}} 
\frac{ V_{\ell_2,m}}{E_m -E_{\ell_2}}
\frac{ V_{m_i',\ell_2}}{V_{m,m} -V_{m_i',m_i'}} 
\right) 
\frac{ V_{n_1,m_i'}}{E_m -E_{n_1}} \right] \right. \nonumber \\
&& \hspace*{1.2cm}\left. 
- \left(
\displaystyle{\sum_{n_2 (\ne m_i)}} 
\frac{ |V_{m,n_2}|^2}{E_m -E_{n_2}}
\right)
\left( 
\displaystyle{\sum_{n_3 (\ne m_i)}} 
\frac{ V_{n_3,m}}{E_m -E_{n_3}}
\frac{ V_{m_i'',n_3}}{V_{m,m} -V_{m_i'',m_i''}} 
\right)
\right\} |m_i'' \rangle, \nonumber \\
\end{eqnarray}

\newpage

where, for example, $V_{m,n}$ is given by $\langle m | V | n \rangle$.

\section{Expressions for Resistivities}
\label{appen_resis}
We describe the expression for $\rho_\sigma^i(\phi)$ 
of Eq. (\ref{rho_sigma^i}) 
up to the second order of 
$\lambda/H$, $\lambda/\Delta$, 
$\lambda/(H \pm \Delta)$, 
$\delta_t/H$, $\delta_t/\Delta$, 
and $\delta_t/(H \pm \Delta)$, 
with $t=\varepsilon$ or $\gamma$. 
Here, we use the following relations: 
\begin{eqnarray}
&& A^2 + B^2 = \frac{1}{\lambda^2}, \\
&& A^2 - B^2 = 
\frac{\delta_\varepsilon}{\lambda^2 \sqrt{\delta_\varepsilon^2 + \lambda^2}}, \\
&& (AB)^2 = \frac{1}{4\lambda^2 (\delta_\varepsilon^2 + \lambda^2)}, \\
&& A^2 \left(\delta_\varepsilon - \sqrt{\delta_\varepsilon^2 + \lambda^2} \right)
+ B^2 \left(\delta_\varepsilon + \sqrt{\delta_\varepsilon^2 + \lambda^2}\right)=0, \\
&& A^2 \left(\delta_\varepsilon - \sqrt{\delta_\varepsilon^2 + \lambda^2}\right)^2
+ B^2 \left(\delta_\varepsilon + \sqrt{\delta_\varepsilon^2 + \lambda^2}\right)^2=1, \\
&& A^4 \left(\delta_\varepsilon - \sqrt{\delta_\varepsilon^2 + \lambda^2}\right)^2
+ B^4 \left(\delta_\varepsilon + \sqrt{\delta_\varepsilon^2 + \lambda^2}\right)^2=
\frac{1}{2 (\delta_\varepsilon^2 + \lambda^2)}, \\
&& A^4 \left(\delta_\varepsilon - \sqrt{\delta_\varepsilon^2 + \lambda^2}\right)^2
- B^4 \left(\delta_\varepsilon + \sqrt{\delta_\varepsilon^2 + \lambda^2}\right)^2=0, \\
&& A^4 + B^4 = \frac{1}{\lambda^4} - \frac{1}{2 \lambda^2  (\delta_\varepsilon^2 + \lambda^2)},
\end{eqnarray}
where $A$ and $B$ are given by Eqs. (\ref{AAA}) and (\ref{BBB}), respectively.

\subsection{${\mbox{\boldmath $I$}}//[100]$}
Using 
Eqs. (\ref{rho_sigma^i})$-$(\ref{tau_sd_inv}), 
we obtain 
$\rho_\pm^{[100]} (\phi)$: 
\begin{eqnarray}
\label{rho_pm^100}
\rho_\pm^{[100]} (\phi) =
\rho_{0,\pm}^{[100]} 
+  \rho_{2,\pm}^{[100]} \cos 2\phi +  \rho_{4,\pm}^{[100]} \cos 4\phi, 
\end{eqnarray}
where $\rho_{0,\pm}^{[100]}$ is a constant term 
independent of $\phi$, 
$\rho_{2,\pm}^{[100]}$ is the coefficient of the $\cos 2\phi$ term, 
and $\rho_{4,\pm}^{[100]}$ is that of the $\cos 4\phi$ term. 
These quantities are specified by 
\begin{eqnarray}
\label{rho_0pm^100}
&&\rho_{0,\pm}^{[100]} = \rho_{0,\pm}^{[100],(0)} + \rho_{0,\pm}^{[100],(2)},  \\
\label{rho_2pm^100}
&&\rho_{2,\pm}^{[100]} = \rho_{2,\pm}^{[100],(1)} + \rho_{2,\pm}^{[100],(2)},  \\
\label{rho_4pm^100}
&&\rho_{4,\pm}^{[100]} = \rho_{4,\pm}^{[100],(2)}, 
\end{eqnarray}
where $v$ of $\rho_{j,\pm}^{[100],(v)}$ ($j=0$, 2, and 4 and $v=0$, 1, and 2) 
denotes the order of 
$\lambda/H$, $\lambda/\Delta$, 
$\lambda/(H \pm \Delta)$, 
$\delta_t/H$, $\delta_t/\Delta$, and 
$\delta_t/(H \pm \Delta)$, 
with $t=\varepsilon$ or $\gamma$. 
The quantity $\rho_{j,\pm}^{[100],(v)}$ was 
given in Eqs. (43), (45), and (46) in Ref. \citen{Kokado3} 
and 
Eqs. (1)$-$(3) in Ref. \citen{Kokado3_1}. 
Note here that 
$\rho_{s,\pm \to m,+}=0$ due to $D_{m,+}^{(d)}=0$ 
and 
Eqs. (\ref{epsilon1}) and (\ref{epsilon2}) 
[i.e., Eqs. (\ref{rho_epsilon1}) and (\ref{rho_epsilon2})] 
are set in Sect. \ref{sec_calc}. 

\subsection{${\mbox{\boldmath $I$}}//[110]$}
Using 
Eqs. (\ref{phi_phi'}) and (\ref{rho_sigma^i})$-$(\ref{tau_sd_inv}), 
we obtain 
$\rho_\pm^{[110]} (\phi')$, 
where 
$\rho_{s,\pm \to m,+}=0$ due to $D_{m,+}^{(d)}=0$ 
is taken into account (see Sect. \ref{sec_calc}). 
Here, we put
\begin{eqnarray}
\label{rho_epsilon1}
&&\rho_{s,\pm \to \delta_\varepsilon,-} \equiv 
\rho_{s,\pm \to \varepsilon 1,-}, \\
\label{rho_epsilon2}
&&\rho_{s,\pm \to \xi_+,-}=
\rho_{s,\pm \to \xi_-,-} \equiv \rho_{s,\pm \to \varepsilon 2,-}, 
\end{eqnarray}
which correspond to Eqs. (\ref{epsilon1}) and (\ref{epsilon2}), respectively 
[also see Eq. (\ref{r_s-d})]. 
It is noted that 
$\rho_\pm^{[110]} (\phi')$ 
composed of 
$\rho_{s,\pm \to \xi_+,-}$ and 
$\rho_{s,\pm \to \xi_-,-}$ 
has very long expressions. 
The expression for $\rho_\pm^{[110]} (\phi')$ is written as
\begin{eqnarray}
\label{rho_pm^110}
\rho_\pm^{[110]} (\phi') =
\rho_{0,\pm}^{[110]} 
+  \rho_{2,\pm}^{[110]} \cos 2\phi' +  \rho_{4,\pm}^{[110]} \cos 4\phi' 
+  \rho_{6,\pm}^{[110]} \cos 6\phi' +  \rho_{8,\pm}^{[110]} \cos 8\phi', 
\end{eqnarray}
where $\rho_{0,\pm}^{[110]}$ is a constant term 
independent of $\phi'$, 
$\rho_{2,\pm}^{[110]}$ is the coefficient of the $\cos 2\phi'$ term, 
$\rho_{4,\pm}^{[110]}$ is that of the $\cos 4\phi'$ term, 
$\rho_{6,\pm}^{[110]}$ is that of the $\cos 6\phi'$ term, 
and $\rho_{8,\pm}^{[110]}$ is that of the $\cos 8\phi'$ term. 
These quantities are specified by 
\begin{eqnarray}
\label{rho_0pm^110}
&&\rho_{0,\pm}^{[110]} = \rho_{0,\pm}^{[110],(0)} + \rho_{0,\pm}^{[110],(2)},  \\
&&\rho_{2,\pm}^{[110]} = \rho_{2,\pm}^{[110],(2)},  \\
\label{rho_4pm^110}
&&\rho_{4,\pm}^{[110]} = \rho_{4,\pm}^{[110],(2)}, \\
&&\rho_{6,+}^{[110]} = 0, \\
&&\rho_{6,-}^{[110]} = \rho_{6,-}^{[110],(2)}, \\
&&\rho_{8,+}^{[110]} = 0, \\
\label{rho_8-^110}
&&\rho_{8,-}^{[110]} = \rho_{8,-}^{[110],(2)}, 
\end{eqnarray}
where $v$ of $\rho_{j,\pm}^{[110],(v)}$ ($j=0$, 2, and 4 and $v=0$, 1, and 2) 
denotes the order of 
$\lambda/H$, $\lambda/\Delta$, 
$\lambda/(H \pm \Delta)$, 
$\delta_t/H$, $\delta_t/\Delta$, and 
$\delta_t/(H \pm \Delta)$, 
with $t=\varepsilon$ or $\gamma$. 
The quantity $\rho_{j,\pm}^{[110],(v)}$ is obtained as
\begin{eqnarray}
\label{rho_0+^110(0)}
&&\rho_{0,+}^{[110],(0)}=\rho_{s,+}, \\
\label{rho_0-^110(0)}
&&\rho_{0,-}^{[110],(0)}=\rho_{s,-}
+\frac{3}{4}\rho_{s,- \to \varepsilon 2,-}
+\frac{1}{4}\rho_{s,- \to 3z^2-r^2,-}, \\
&&\rho_{0,+}^{[110],(2)}=
\frac{3}{32}
\left( \frac{\lambda}{H- \Delta} \right)^2 
(\rho_{s,+ \to \varepsilon 1, -} + \rho_{s,+ \to \varepsilon 2, -} )
+ \frac{3}{16} 
\left( \frac{\lambda}{H} \right)^2 
\rho_{s,+ \to \varepsilon 1,-} \nonumber \\
&&\hspace*{1.8cm}
+ \frac{3}{4}
\left( \frac{\lambda}{H+ \Delta} \right)^2 
\rho_{s,+ \to x^2-y^2,-}, \\
&&\rho_{0,-}^{[110],(2)}=
-\frac{3}{16}
\left( \frac{\lambda}{H} \right)^2 
\rho_{s,- \to \varepsilon 2,-} 
-\frac{3}{4}
\left( \frac{\lambda}{H-\Delta} \right)^2 
\left( \rho_{s,- \to \varepsilon 2,-}
+ \frac{1}{4} \rho_{s,- \to 3z^2-r^2,-} \right) \nonumber \\
&&\hspace*{1.9cm}+\frac{27}{128}
\left( \frac{\lambda}{H-\Delta} + \frac{\lambda}{\Delta} \right)^2 
( \rho_{s,- \to \varepsilon 1,-}
- \rho_{s,- \to \varepsilon 2,-} ) \nonumber \\
&&\hspace*{1.9cm}+\frac{3}{32}
\left(\frac{\lambda}{\Delta} \right)^2 
( \rho_{s,- \to \varepsilon 1,-}
+ \rho_{s,- \to \varepsilon 2,-} 
-2 \rho_{s,- \to 3z^2-r^2,-}) \nonumber \\
&&\hspace*{1.9cm}+\frac{3}{128}
\left( \frac{\lambda}{\delta_\gamma} \right)^2 
\left( \frac{\lambda}{H+\Delta} - \frac{\lambda}{\Delta} \right)^2 
( \rho_{s,- \to x^2-y^2,-}
- \rho_{s,- \to 3z^2-r^2,-}), \\
\label{rho_2+^110(2)}
&&\rho_{2,+}^{[110],(2)}=
\frac{3}{8} \frac{\lambda^2}{H(H-\Delta)}
\rho_{s,+ \to \varepsilon 1,-}, \\
\label{rho_2-^110(2)}
&&\rho_{2,-}^{[110],(2)}=
\frac{9}{32} \frac{\lambda}{\Delta}
\left( \frac{\lambda}{H-\Delta} + \frac{\lambda}{\Delta} \right)
(\rho_{s,- \to \varepsilon 1,-} -\rho_{s,- \to \varepsilon 2,-}) 
\nonumber \\
&& \hspace*{1.9cm} 
+ \frac{3}{8}
\left( \frac{\lambda}{\Delta} \right)^2 
(\rho_{s,- \to \varepsilon 2,-} -\rho_{s,- \to 3z^2-r^2,-}) \nonumber \\
&&\hspace*{1.9cm}-\frac{3}{8} \frac{\lambda^2}{H \Delta} \rho_{s,- \to \varepsilon 2,-}
+\frac{3}{8} \frac{\lambda^2}{\Delta (H+\Delta)} \rho_{s,- \to 3z^2-r^2,-}, \\
\label{rho_4+^110(2)}
&&\rho_{4,+}^{[110],(2)}=\frac{3}{32}\left( \frac{\lambda}{H-\Delta}\right)^2 
(\rho_{s,+ \to \varepsilon 1,-} -\rho_{s,+ \to \varepsilon 2,-}),  \\
\label{rho_4-^110(2)}
&&\rho_{4,-}^{[110],(2)}=
\frac{3}{32} \left( \frac{\lambda}{\Delta} \right)^2 
(\rho_{s,- \to \varepsilon 2,-} -\rho_{s,- \to \varepsilon 1,-}) \nonumber \\
&&\hspace*{1.9cm}+ \frac{3}{128} 
\left( \frac{\lambda}{\delta_\gamma} \right)^2 
\left( \frac{\lambda}{H+\Delta} - \frac{\lambda}{\Delta} \right)^2 
(\rho_{s,- \to 3z^2-r^2,-} -\rho_{s,- \to x^2-y^2,-}), \\
\label{rho_6-^110(2)}
&&\rho_{6,-}^{[110],(2)}=
\frac{9}{32} \frac{\lambda}{\Delta}
\left( \frac{\lambda}{H-\Delta} + \frac{\lambda}{\Delta} \right)
(\rho_{s,- \to \varepsilon 2,-} -\rho_{s,- \to \varepsilon 1,-}), \\
\label{rho_8-^110(2)}
&&\rho_{8,-}^{[110],(2)}=
\frac{27}{128} 
\left( \frac{\lambda}{H-\Delta} + \frac{\lambda}{\Delta} \right)^2
(\rho_{s,- \to \varepsilon 2,-} -\rho_{s,- \to \varepsilon 1,-}).
\end{eqnarray}

\subsection{${\mbox{\boldmath $I$}}//[001]$}
Using 
Eqs. (\ref{rho_sigma^i})$-$(\ref{tau_sd_inv}), 
we obtain 
$\rho_\pm^{[001]} (\phi)$, 
where 
$\rho_{s,\pm \to m,+}=0$ due to $D_{m,+}^{(d)}=0$ 
is taken into account (see Sect. \ref{sec_calc}). 
The resistivity $\rho_\pm^{[001]} (\phi)$ is written as
\begin{eqnarray}
\label{rho_pm^001}
\rho_\pm^{[001]} (\phi) =
\rho_{0,\pm}^{[001]} 
+  \rho_{4,\pm}^{[001]} \cos 4\phi, 
\end{eqnarray}
where $\rho_{0,\pm}^{[001]}$ is a constant term 
independent of $\phi$, 
and $\rho_{4,\pm}^{[001]}$ is the coefficient of the $\cos 4\phi$ term. 
These quantities are specified by 
\begin{eqnarray}
\label{rho_0pm^001}
&&\rho_{0,\pm}^{[001]} = \rho_{0,\pm}^{[001],(0)} + \rho_{0,\pm}^{[001],(2)},  \\
\label{rho_4pm^001}
&&\rho_{4,\pm}^{[001]} = \rho_{4,\pm}^{[001],(2)}, 
\end{eqnarray}
where $v$ of $\rho_{j,\pm}^{[100],(v)}$ ($j=0$ and 4 and $v=0$, 1, and 2) 
denotes the order of 
$\lambda/H$, $\lambda/\Delta$, 
$\lambda/(H \pm \Delta)$, 
$\delta_t/H$, $\delta_t/\Delta$, and 
$\delta_t/(H \pm \Delta)$, 
with $t=\varepsilon$ or $\gamma$. 
The quantity $\rho_{j,\pm}^{[001],(v)}$ is obtained as
\begin{eqnarray}
\label{rho_0+^001(0)}
&&\rho_{0,+}^{[001],(0)}=\rho_{s,+}, \\
\label{rho_0-^001(0)}
&&\rho_{0,-}^{[001],(0)} =\rho_{s,-}+\rho_{s,- \to 3z^2-r^2,-}, \\
&&\rho_{0,+}^{[001],(2)} =
\frac{3}{8} \left( \frac{ \lambda}{H-\Delta} \right)^2 
\left( \lambda^2 A^2 \rho_{s,+ \to \xi_+,-}
+
\rho_{s,+ \to \delta_\varepsilon,-}
+
\lambda^2 B^2 \rho_{s,+ \to \xi_-,-} \right), \\
&&\rho_{0,-}^{[001],(2)} =
\frac{3}{8}
\left( \frac{\lambda}{\Delta} \right)^2 
( \lambda^2 A^2 \rho_{s,- \to \xi_+,-} + 
\rho_{s,- \to \delta_\varepsilon,-}
+ \lambda^2 B^2 \rho_{s,- \to \xi_-,-} )\nonumber \\
&&\hspace*{1.9cm}
+ \frac{3}{32} 
\left( \frac{\lambda}{\delta_\gamma} \right)^2 
\left( \frac{\lambda}{H+\Delta} - \frac{\lambda}{\Delta} \right)^2 
\rho_{s,- \to x^2-y^2,-} \nonumber \\
&&\hspace*{1.9cm} - \frac{3}{4} \left[ 
\left( \frac{\lambda}{H+\Delta} \right)^2 
+ \left( \frac{\lambda}{\Delta} \right)^2 
+ \frac{1}{8} 
\left( \frac{\lambda}{\delta_\gamma} \right)^2 
\left( \frac{\lambda}{H+\Delta} - \frac{\lambda}{\Delta} \right)^2 
\right] \rho_{s,- \to 3z^2-r^2,-}, \nonumber \\ \\
\label{rho_4+^001(2)}
&&\rho_{4,+}^{[001],(2)} =
\frac{3}{8} \left( \frac{ \lambda }{H-\Delta} \right)^2
\left( \lambda^2 A^2 \rho_{s,+ \to \xi_+,-}
-
\rho_{s,+ \to \delta_\varepsilon,-}
+
\lambda^2 B^2 \rho_{s,+ \to \xi_-,-} \right), \\
\label{rho_4-^001(2)}
&&\rho_{4,-}^{[001],(2)} =
\frac{3}{8} \left( \frac{\lambda}{\Delta} \right)^2 
\left( 
-\lambda^2 A^2 \rho_{s,- \to \xi_+,-} 
+ \rho_{s,- \to \delta_\varepsilon,-} 
- \lambda^2 B^2 \rho_{s,- \to \xi_-,-} \right) \nonumber \\
&&\hspace*{1.9cm}+ \frac{3}{32} 
\left( \frac{\lambda}{\delta_\gamma} \right)^2 
\left( \frac{\lambda}{H+\Delta} - \frac{\lambda}{\Delta} \right)^2 
(\rho_{s,- \to x^2-y^2,-} - \rho_{s,- \to 3z^2-r^2,-} ). 
\end{eqnarray}
Note that Eqs. (\ref{rho_epsilon1}) and (\ref{rho_epsilon2}) 
[i.e., Eqs. (\ref{epsilon1}) and (\ref{epsilon2})] 
are introduced 
in Sect. \ref{sec_cons}.

\section{Coefficient Expressed by Using Resistivities} 
\label{appen_coeff}
We express 
$C_j^i$ as a function of $\rho_{j,\sigma}^{i,(v)}$. 
Here, $C_j^i$ is expressed 
up to the second order of 
$\lambda/H$, $\lambda/\Delta$, 
$\lambda/(H \pm \Delta)$, 
$\delta_t/H$, $\delta_t/\Delta$, 
and $\delta_t/(H \pm \Delta)$, 
with $t=\varepsilon$ or $\gamma$.

\subsection{${\mbox{\boldmath $I$}}//[100]$}
Using Eqs. (\ref{AMR^i(phi_i)}), (\ref{rho^i(phi)}), 
(\ref{rho_pm^100})$-$(\ref{rho_4pm^100}), 
we obtain 
$C_2^{[100]}$ and $C_4^{[100]}$ in AMR$^{[100]}(\phi)$ 
of Eq. (\ref{AMR^100}): 
\begin{eqnarray}
\label{C_2^100_rho}
&&C_2^{[100]}=-\frac{\rho_{2,+}^{[100],(1)} + \rho_{2,-}^{[100],(1)}}
{( \rho_{0,+}^{[100],(0)} + \rho_{0,-}^{[100],(0)} )^2}
\left( \frac{\rho_{0,+}^{[100],(0)}}{\rho_{0,-}^{[100],(0)}}
\rho_{2,-}^{[100],(1)}
+ \frac{\rho_{0,-}^{[100],(0)}}{\rho_{0,+}^{[100],(0)}}
\rho_{2,+}^{[100],(1)}\right) \nonumber \\
&&\hspace*{1.5cm}+ 
\frac{1}
{\rho_{0,+}^{[100],(0)} + \rho_{0,-}^{[100],(0)}}
\left( \frac{\rho_{0,+}^{[100],(0)}}{\rho_{0,-}^{[100],(0)}}
\rho_{2,-}^{[100],(2)}
+ \frac{\rho_{0,-}^{[100],(0)}}{\rho_{0,+}^{[100],(0)}}
\rho_{2,+}^{[100],(2)}\right) \nonumber \\
&&\hspace*{1.5cm}+
\frac{1}
{\rho_{0,+}^{[100],(0)} + \rho_{0,-}^{[100],(0)}}
\left( \frac{\rho_{2,-}^{[100],(1)}}{\rho_{0,-}^{[100],(0)}}
+ \frac{\rho_{2,+}^{[100],(1)}}{\rho_{0,+}^{(0)}} \right)
\left( \frac{\rho_{0,+}^{[100],(0)}}{\rho_{0,-}^{[100],(0)}}
\rho_{2,-}^{[100],(1)}
+ \frac{\rho_{0,-}^{[100],(0)}}{\rho_{0,+}^{[100],(0)}}
\rho_{2,+}^{[100],(1)}\right) \nonumber \\
&&\hspace*{1.5cm} + 
\frac{1}
{\rho_{0,+}^{[100],(0)} + \rho_{0,-}^{[100],(0)}}
\left( \frac{\rho_{0,+}^{[100],(0)}}{\rho_{0,-}^{[100],(0)}}
\rho_{2,-}^{[100],(1)}
+ \frac{\rho_{0,-}^{[100],(0)}}{\rho_{0,+}^{[100],(0)}}
\rho_{2,+}^{[100],(1)}\right),
\\
\label{C_4^100_rho}
&&C_4^{[100]}=
\frac{1}
{\rho_{0,+}^{[100],(0)} + \rho_{0,-}^{[100],(0)}}
\left( \frac{\rho_{0,+}^{[100],(0)}}{\rho_{0,-}^{[100],(0)}}
\rho_{4,-}^{[100],(2)}
+ \frac{\rho_{0,-}^{[100],(0)}}{\rho_{0,+}^{[100],(0)}}
\rho_{4,+}^{[100],(2)}\right)
+
\frac{1}{2}
\frac{\rho_{2,+}^{[100],(1)} \rho_{2,-}^{[100],(1)}}
{\rho_{0,+}^{[100],(0)} \rho_{0,-}^{[100],(0)}} \nonumber \\
&&\hspace*{1.5cm}
-\frac{1}{2}\frac{\rho_{2,+}^{[100],(1)} + \rho_{2,-}^{[100],(1)}}
{( \rho_{0,+}^{[100],(0)} + \rho_{0,-}^{[100],(0)} )^2}
\left( \frac{\rho_{0,+}^{[100],(0)}}{\rho_{0,-}^{[100],(0)}}
\rho_{2,-}^{(1)}
+ \frac{\rho_{0,-}^{[100],(0)}}{\rho_{0,+}^{[100],(0)}}
\rho_{2,+}^{[100],(1)}\right).
\end{eqnarray}
In the case of $\rho_{s,\pm \to x^2-y^2,\pm} = \rho_{s,\pm \to 3z^2-r^2,\pm}$, 
we have $\rho_{2,\pm}^{[100],(1)}=0$. 
In this case, $C_2^{[100]}$ and $C_4^{[100]}$ become
\begin{eqnarray}
\label{C_2^100_gamma}
&&C_2^{[100]}=
\frac{1}
{\rho_{0,+}^{[100],(0)} + \rho_{0,-}^{[100],(0)}}
\left( \frac{\rho_{0,+}^{[100],(0)}}{\rho_{0,-}^{[100],(0)}}
\rho_{2,-}^{[100],(2)}
+ \frac{\rho_{0,-}^{[100],(0)}}{\rho_{0,+}^{[100],(0)}}
\rho_{2,+}^{[100],(2)}\right),
\\
\label{C_4^100_gamma}
&&C_4^{[100]}=
\frac{1}
{\rho_{0,+}^{[100],(0)} + \rho_{0,-}^{[100],(0)}}
\left( \frac{\rho_{0,+}^{[100],(0)}}{\rho_{0,-}^{[100],(0)}}
\rho_{4,-}^{[100],(2)}
+ \frac{\rho_{0,-}^{[100],(0)}}{\rho_{0,+}^{[100],(0)}}
\rho_{4,+}^{[100],(2)}\right).
\end{eqnarray}
Note here that 
$\cos j\phi$ of $C_j^{[100]} \cos j\phi$ comes from 
$\cos j\phi$ of $\rho_{j,\pm}^{[100],(2)} \cos j\phi$, where $j=2$ and 4.

\subsection{${\mbox{\boldmath $I$}}//[110]$}
Using Eqs. (\ref{AMR^i(phi_i)}), (\ref{rho^i(phi)}), 
and (\ref{rho_pm^110})$-$(\ref{rho_8-^110}), 
we obtain 
$C_2^{[110]}$, $C_4^{[110]}$, $C_6^{[110]}$, and $C_8^{[110]}$ 
in AMR$^{[110]}(\phi')$ of Eq. (\ref{AMR^110}): 
\begin{eqnarray}
\label{C_2^110_rho}
&&C_2^{[110]}=
\frac{1}
{\rho_{0,+}^{[110],(0)} + \rho_{0,-}^{[110],(0)}}
\left( \frac{\rho_{0,+}^{[110],(0)}}{\rho_{0,-}^{[110],(0)}}
\rho_{2,-}^{[110],(2)}
+ \frac{\rho_{0,-}^{[110],(0)}}{\rho_{0,+}^{[110],(0)}}
\rho_{2,+}^{[110],(2)}\right),
\\
\label{C_4^110_rho}
&&C_4^{[110]}=
\frac{1}
{\rho_{0,+}^{[110],(0)} + \rho_{0,-}^{[110],(0)}}
\left( \frac{\rho_{0,+}^{[110],(0)}}{\rho_{0,-}^{[110],(0)}}
\rho_{4,-}^{[110],(2)}
+ \frac{\rho_{0,-}^{[110],(0)}}{\rho_{0,+}^{[110],(0)}}
\rho_{4,+}^{[110],(2)}\right), \\
\label{C_6^110_rho}
&&C_6^{[110]}
= \frac{\rho_{0,+}^{[110],(0)} \rho_{6,-}^{[110],(2)}}
{\rho_{0,-}^{[110],(0)}
(\rho_{0,+}^{[110],(0)} + \rho_{0,-}^{[110],(0)})}, \\
\label{C_8^110_rho}
&&C_8^{[110]}
= \frac{\rho_{0,+}^{[110],(0)} \rho_{8,-}^{[110],(2)}}
{\rho_{0,-}^{[110],(0)}
(\rho_{0,+}^{[110],(0)} + \rho_{0,-}^{[110],(0)})}.
\end{eqnarray}
Note here that 
$\cos j\phi$ of $C_j^{[110]} \cos j\phi$ comes from 
$\cos j\phi$ of $\rho_{j,\pm}^{[110],(2)} \cos j\phi$, 
where $j=2$ and 4. 
In addition, 
$\cos j'\phi$ of $C_{j'}^{[110]} \cos j'\phi$ comes from 
$\cos j'\phi$ of $\rho_{j',-}^{[110],(2)} \cos j'\phi$, 
where $j'=6$ and 8.

\subsection{${\mbox{\boldmath $I$}}//[001]$}
Using Eqs. (\ref{AMR^i(phi_i)}), (\ref{rho^i(phi)}), 
and (\ref{rho_pm^001})$-$(\ref{rho_4pm^001}), 
we obtain 
$C_4^{[001]}$ in AMR$^{[001]}(\phi)$ 
of Eq. (\ref{AMR^001}): 
\begin{eqnarray}
\label{C_4^001_rho}
&&C_4^{[001]}=
\frac{1}
{\rho_{0,+}^{[001],(0)} + \rho_{0,-}^{[001],(0)}}
\left( \frac{\rho_{0,+}^{[001],(0)}}{\rho_{0,-}^{[001],(0)}}
\rho_{4,-}^{[001],(2)}
+ \frac{\rho_{0,-}^{[001],(0)}}{\rho_{0,+}^{[001],(0)}}
\rho_{4,+}^{[001],(2)}\right). 
\end{eqnarray}
Note here that 
$\cos 4\phi$ of $C_4^{[001]} \cos 4\phi$ comes from 
$\cos 4\phi$ of $\rho_{4,\pm}^{[001],(2)} \cos 4\phi$.

\section{Expressions for $C_j^i$
} 
\label{appen_C_j^i}

We give expressions for $C_j^i$ in 
Sects. \ref{sub_I//100}, \ref{sub_I//110}, and \ref{sub_I//001}. 

\subsection{${\mbox{\boldmath $I$}}//[100]$} 
\label{appen_C_j^100}
The expressions for $C_0^{[100]}$, $C_2^{[100]}$, and $C_4^{[100]}$ 
are 
\begin{eqnarray}
\label{C_0^100}
&&C_0^{[100]} = C_2^{[100]} - C_4^{[100]}, \\
\label{C_2^100}
&&C_2^{[100]} =
\frac{3}{8} 
\frac{1}{ 1+r+ \frac{3}{4}r_{s,- \to x^2-y^2, -}+ \frac{1}{4}r_{s,- \to 3z^2-r^2, -}} \nonumber \\
&& \hspace*{1.5cm} \times 
\Bigg\{ 
\frac{1}{r+ \frac{3}{4}r_{s,- \to x^2-y^2, -}+ \frac{1}{4}r_{s,- \to 3z^2-r^2, -}} 
\Bigg\{ 
- \left( \frac{\lambda}{\Delta} \right)^2 
r_{s,- \to \varepsilon 1, -}   \nonumber \\
&& \hspace*{1.5cm} 
+ \left[
\left( \frac{\lambda}{\Delta} \right)^2 
- \left( \frac{\lambda}{H+\Delta} \right)^2 
\right] 
r_{s,- \to 3z^2-r^2, -}  \nonumber \\
&&  \hspace*{1.5cm} 
+ \frac{\lambda}{\delta_\gamma} 
\left[ 
\frac{\lambda \delta_\varepsilon}{\Delta^2} 
- \left( \frac{\lambda}{H+\Delta} \right)^2 
\left( \frac{\delta_\varepsilon}{\lambda} -1 \right)
-  \frac{\lambda^2}{\Delta (H+\Delta)} \right] 
(r_{s,- \to x^2-y^2, -} - r_{s,- \to 3z^2-r^2, -})   \nonumber \\
&& \hspace*{1.5cm}  - \frac{1}{2} 
\left( \frac{\lambda}{\delta_\gamma} \right)^2 
\left( 
\frac{\lambda}{\Delta} 
- \frac{\lambda}{H+\Delta} \right)^2 
(r_{s,- \to x^2-y^2, -} - r_{s,- \to 3z^2-r^2, -})\Bigg\}  \nonumber \\
&& \hspace*{1.5cm}  + \left( \frac{\lambda}{H-\Delta} \right)^2 
r_{s,+ \to \varepsilon 2, -}
\left( 
r+ \frac{3}{4}r_{s,- \to x^2-y^2, -}+ \frac{1}{4}r_{s,- \to 3z^2-r^2, -} 
\right) \nonumber \\
&& \hspace*{1.5cm}  +
\frac{\lambda}{\delta_\gamma}
\left( 
\frac{\lambda}{\Delta} 
- \frac{\lambda}{H+\Delta} \right) 
\frac{r_{s,- \to x^2-y^2, -} - r_{s,- \to 3z^2-r^2, -}}{r+ \frac{3}{4}r_{s,- \to x^2-y^2, -}+ \frac{1}{4}r_{s,- \to 3z^2-r^2, -}}  \nonumber \\
&& \hspace*{1.5cm}  + \frac{3}{8}
\left( \frac{\lambda}{\delta_\gamma} \right)^2 
\left( 
\frac{\lambda}{\Delta} 
- \frac{\lambda}{H+\Delta} \right)^2 
\frac{(r_{s,- \to x^2-y^2, -} - r_{s,- \to 3z^2-r^2, -})^2}
{r+ \frac{3}{4}r_{s,- \to x^2-y^2, -}+ \frac{1}{4}r_{s,- \to 3z^2-r^2, -}}
 \nonumber \\
&& \hspace*{1.5cm}  \times \left[ 
\frac{1}{1+r+ \frac{3}{4}r_{s,- \to x^2-y^2, -}+ \frac{1}{4}r_{s,- \to 3z^2-r^2, -}} + \frac{1}
{r+ \frac{3}{4}r_{s,- \to x^2-y^2, -}+ \frac{1}{4}r_{s,- \to 3z^2-r^2, -}} 
\right]
\Bigg\}, \nonumber \\\\
\label{C_4^100}
&&C_4^{[100]} =
\frac{3}{32}
\frac{1}{ 1+r+ \frac{3}{4}r_{s,- \to x^2-y^2, -}+ \frac{1}{4}r_{s,- \to 3z^2-r^2, -}} \nonumber \\
&& \hspace*{1.5cm} \times 
\Bigg\{ 
\frac{1}{r+ \frac{3}{4}r_{s,- \to x^2-y^2, -}+ \frac{1}{4}r_{s,- \to 3z^2-r^2, -}} 
\left[ 
\left( \frac{\lambda}{\Delta} \right)^2 
(r_{s,- \to \varepsilon 1, -} - r_{s,- \to \varepsilon 2, -}) \right.  \nonumber \\
&& \hspace*{1.5cm}  \left. + 
\frac{1}{2}
\left( \frac{\lambda}{\delta_\gamma} \right)^2 
\left( 
\frac{\lambda}{\Delta} 
- \frac{\lambda}{H+\Delta} \right)^2 
(r_{s,- \to 3z^2-r^2, -} - r_{s,- \to x^2-y^2, -})
\right]  \nonumber \\
&& \hspace*{1.5cm} + 
\left( \frac{\lambda}{H-\Delta} \right)^2 
(r_{s,+ \to \varepsilon 2, -} - r_{s,+ \to \varepsilon 1, -})
\left( 
r+ \frac{3}{4}r_{s,- \to x^2-y^2, -}+ \frac{1}{4}r_{s,- \to 3z^2-r^2, -}
\right)
 \nonumber \\
&& \hspace*{1.5cm} 
- \frac{3}{4}
\left( \frac{\lambda}{\delta_\gamma} \right)^2 
\left( 
\frac{\lambda}{\Delta} 
- \frac{\lambda}{H+\Delta} \right)^2  \nonumber \\
&& \hspace*{1.5cm} 
\times \frac{(r_{s,- \to 3z^2-r^2, -} - r_{s,- \to x^2-y^2, -})^2}
{(1+r+ \frac{3}{4}r_{s,- \to x^2-y^2, -}+ \frac{1}{4}r_{s,- \to 3z^2-r^2, -})
(r+ \frac{3}{4}r_{s,- \to x^2-y^2, -}+ \frac{1}{4}r_{s,- \to 3z^2-r^2, -})
}
\Bigg\}. \nonumber \\
\end{eqnarray}

\subsection{${\mbox{\boldmath $I$}}//[110]$} 
\label{appen_C_j^110}
The expressions for $C_0^{[110]}$, $C_2^{[110]}$, $C_4^{[110]}$, 
$C_6^{[110]}$, and $C_8^{[110]}$ are 
\begin{eqnarray}
\label{C_0^110}
&&\hspace*{-0.8cm}C_0^{[110]}=C_2^{[110]} - C_4^{[110]} + C_6^{[110]} - C_8^{[110]}, \\
\label{C_2^110}
&&\hspace*{-0.8cm}C_2^{[110]} = 
\frac{3}{8}
\frac{1}
{\left( r + \displaystyle{\frac{3}{4}} r_{s,- \to \varepsilon 2, -} 
+\displaystyle{\frac{1}{4}} r_{s,- \to 3z^2-r^2, -} \right)
\left( 1 + r + \displaystyle{\frac{3}{4}} r_{s,- \to \varepsilon 2, -}
+ \displaystyle{\frac{1}{4}} r_{s,- \to 3z^2-r^2, -}
\right)} \nonumber \\ 
&& \hspace*{0.6cm}
\times \left[ 
\left( \frac{\lambda}{\Delta} \right)^2
(r_{s,- \to \varepsilon 2, -} - 
r_{s,- \to 3z^2-r^2, -} )
+ \displaystyle{\frac{3}{4}} 
\frac{\lambda}{\Delta} \left( \frac{\lambda}{H-\Delta} + \frac{\lambda}{\Delta} \right)
(r_{s,- \to \varepsilon 1, -} - 
r_{s,- \to \varepsilon 2, -} ) \right. \nonumber \\
&&\hspace*{0.7cm}\left. - \frac{\lambda^2}{H \Delta}
r_{s,- \to \varepsilon 2, -} 
+ \frac{\lambda^2}{\Delta (H +\Delta)}
r_{s,- \to 3z^2-r^2, -} \right] \nonumber \\
&&\hspace*{0.7cm} +
\frac{3}{8}\frac{\lambda^2}{H(H-\Delta)} 
\frac{
 r_{s,+ \to \varepsilon 1, -}
\left( r + \displaystyle{\frac{3}{4}}  r_{s,- \to \varepsilon 2, -}
+  \displaystyle{\frac{1}{4}}r_{s,- \to 3z^2-r^2, -} \right)}
{1 + r + \displaystyle{\frac{3}{4}}  r_{s,- \to \varepsilon 2, -}
+ \frac{1}{4}  r_{s,- \to 3z^2-r^2, -} } , \\
\label{C_4^110}
&&\hspace*{-0.8cm}C_4^{[110]} = 
\frac{3}{32}
\frac{1}
{ \left( r + \displaystyle{\frac{3}{4}} r_{s,- \to \varepsilon 2, -}
+ \displaystyle{\frac{1}{4}} r_{s,- \to 3z^2-r^2, -} \right)
\left(1 + r + \displaystyle{\frac{3}{4}} r_{s,- \to \varepsilon 2, -} 
+ \displaystyle{\frac{1}{4}} r_{s,- \to 3z^2-r^2, -} \right)} \nonumber \\
&&\hspace*{0.7cm}
\times 
\Bigg[ 
\left( \frac{\lambda }{\Delta} \right)^2
(r_{s,- \to \varepsilon 2, -}
-r_{s,- \to \varepsilon 1, -} )\nonumber \\
&&\hspace*{0.7cm}
+\frac{1}{4} \left( \frac{\lambda }{\delta_\gamma} \right)^2
\left( \frac{\lambda}{H+\Delta} - \frac{\lambda}{\Delta} \right)^2 
(
r_{s,- \to 3z^2-r^2, -}
-r_{s,- \to x^2-y^2, -} )
\Bigg]
\nonumber \\
&&\hspace*{0.7cm}+
\frac{3}{32} \left( \frac{\lambda }{H-\Delta} \right)^2
\frac{
(r_{s,+ \to \varepsilon 1, -} - r_{s,+ \to \varepsilon 2, -})
\left( r + \displaystyle{\frac{3}{4}}r_{s,- \to \varepsilon 2, -} 
+ \displaystyle{\frac{1}{4}} r_{s,- \to 3z^2-r^2, -} \right)}
{1 + r + \displaystyle{\frac{3}{4}} r_{s,- \to \varepsilon 2, -} 
+ \displaystyle{\frac{1}{4}} r_{s,- \to 3z^2-r^2, -}}, \\
&&\hspace*{-0.8cm}C_6^{[110]} =
\frac{9}{32} \left( \frac{\lambda }{\Delta} \right) \left( \frac{\lambda}{H-\Delta} + \frac{\lambda}{\Delta} \right) \nonumber \\
&& \hspace*{0.7cm} \times\frac{r_{s,- \to \varepsilon 2, -} - r_{s,- \to \varepsilon 1, -} }
{\left(r + \displaystyle{\frac{3}{4}} r_{s,- \to \varepsilon 2, -}
+ \frac{1}{4} r_{s,- \to 3z^2-r^2, -} \right)
\left(1 + r + \displaystyle{\frac{3}{4}} r_{s,- \to \varepsilon 2, -} 
+ \frac{1}{4} r_{s,- \to 3z^2-r^2, -} \right)}, \\
\label{C_8^110}
&&\hspace*{-0.8cm}C_8^{[110]} =
\frac{27}{128} \left( \frac{\lambda}{H-\Delta} + \frac{\lambda}{\Delta} \right) ^2 \nonumber \\
&& \hspace*{0.7cm} \times \frac{r_{s,- \to \varepsilon 2, -} - r_{s,- \to \varepsilon 1, -} }
{\left( r + \displaystyle{\frac{3}{4}} r_{s,- \to \varepsilon 2, -}
+ \frac{1}{4} r_{s,- \to 3z^2-r^2, -} \right)
\left( 1 + r + \displaystyle{\frac{3}{4}} r_{s,- \to \varepsilon 2, -} 
+ \frac{1}{4} r_{s,- \to 3z^2-r^2, -} \right)}. 
\end{eqnarray}
Here, we used 
$\sin 2\phi=\cos 2\phi'$, 
$\cos 2\phi=-\sin 2\phi'$, 
$\sin 4\phi=-\sin 4\phi'$, 
$\cos 4\phi=-\cos 4\phi'$, 
$\sin 6\phi=-\cos 6\phi'$, 
and $\cos 8\phi=\cos 8\phi'$, 
where 
the relation between $\phi$ and $\phi'$ is 
given by Eq. (\ref{phi_phi'}). 

\subsection{${\mbox{\boldmath $I$}}//[001]$} 
\label{appen_C_j^001}
The expressions for $C_0^{[001]}$ and $C_4^{[001]}$ are 
\begin{eqnarray}
\label{C_0^001}
&&C_0^{[001]}=- C_4^{[001]}, \\
\label{C_4^001}
&&C_4^{[001]} =
\frac{3}{8}
\frac{1}{1 + r + r_{s,- \to 3z^2-r^2, -}} 
\Bigg[
\left( \frac{\lambda }{\Delta} \right)^2 
\frac{ r_{s,- \to \varepsilon 1, -} - r_{s,- \to \varepsilon 2, -}}
{r + r_{s,- \to 3z^2-r^2, -}} \nonumber \\
&&\hspace*{1.5cm}+ 
\frac{1}{4}
\left( \frac{\lambda }{\delta_\gamma} \right)^2 
\left( \frac{\lambda}{\Delta} - \frac{\lambda}{H+\Delta} \right)^2 
\frac{ r_{s,- \to x^2-y^2, -} - r_{s,- \to 3z^2-r^2, -}}
{r + r_{s,- \to 3z^2-r^2, -}}  \nonumber \\
&&\hspace*{1.5cm} + 
\left( \frac{\lambda }{H-\Delta} \right)^2 
(r_{s,+ \to \varepsilon 2, -} - r_{s,+ \to \varepsilon 1, -})
(r + r_{s,- \to 3z^2-r^2, -}) \Bigg]. 
\end{eqnarray}

\section{Origin of $C_j^i \cos j \phi_i$}
\label{appen_origin}
We explain the origin of $C_j^i \cos j \phi_i$. 

\subsection{${\mbox{\boldmath $I$}}//[100]$}
As shown in Table \ref{table4}, 
$C_2^{[100]} \cos 2\phi$ 
is related to the probability amplitudes of 
$|3z^2 -r^2, \chi_\pm (\phi) \rangle$ 
and $|x^2 -y^2, \chi_- (\phi) \rangle$, 
and 
$C_4^{[100]} \cos 4\phi$ 
is related to the probability of 
$|3z^2 -r^2, \chi_\pm (\phi) \rangle$.\cite{fig} 

\subsection{${\mbox{\boldmath $I$}}//[110]$} 
As shown in Table \ref{table4}, 
$C_2^{[110]} \cos 2\phi'$ is related to 
the probability amplitude of 
$|3z^2 -r^2, \chi_\pm (\phi) \rangle$, 
the probability amplitude of 
$|xy, \chi_- (\phi) \rangle$, 
and 
the product of the probability amplitude of 
$|3z^2 -r^2, \chi_- (\phi) \rangle$ 
and the probability amplitude of 
$|xy,\chi_- (\phi) \rangle$. 
The term $C_4^{[110]} \cos 4\phi'$ 
is related to the probability of 
$|3z^2 -r^2, \chi_\pm (\phi) \rangle$. 
The term $C_6^{[110]} \cos 6\phi'$ 
is related to 
the probability amplitude of 
$|3z^2 -r^2, \chi_- (\phi) \rangle$ 
and 
the product of the probability amplitude of 
$|3z^2 -r^2, \chi_- (\phi) \rangle$ 
and the probability amplitude of 
$|xy,\chi_- (\phi) \rangle$. 
The term 
$C_8^{[110]} \cos 8\phi'$ 
is related to the probability of 
$|xy, \chi_- (\phi) \rangle$ 
and the probability amplitude of 
$|xy, \chi_- (\phi) \rangle$. 

\subsection{${\mbox{\boldmath $I$}}//[001]$}
As shown in Table \ref{table4}, 
$C_4^{[001]} \cos 4\phi$ 
is related to the probability of 
$|3z^2 -r^2, \chi_\pm (\phi) \rangle$. 

\section{Correspondence to Campbell--Fert--Jaoul Model}
\label{appen_CFJ}
We confirm that 
AMR$^{[100]}(0)$ and AMR$^{[110]}(0)$
correspond to 
the AMR ratio of the CFJ model\cite{Campbell} 
under the condition of the CFJ model, i.e., 
$\rho_{s, \sigma \to m,-}/\rho_{s,+} \equiv \alpha$, 
$r \ll 1$, and $r \ll \alpha$.\cite{Kokado1,[001]} 
Here, we take into account $\Delta/H \ll 1$.

\begin{enumerate}
\item[(1)] ${\mbox{\boldmath $I$}}//[100]$ \\
Under the condition of the CFJ model, 
$C_2^{[100]}$ in Table \ref{table1} becomes
\begin{eqnarray}
\label{C2_100_CFJ}
&&C_2^{[100]}= \frac{3}{8} \frac{1}{1+r+\alpha} 
\left[ - \left( \frac{\lambda}{H} \right)^2 
\frac{\alpha}{r+\alpha} + 
\left( \frac{\lambda}{H} \right)^2 
\alpha (r + \alpha) \right] \nonumber \\
&& \hspace*{1.1cm}\approx \frac{3}{8} \left( \frac{\lambda}{H} \right)^2 
(\alpha -1). 
\end{eqnarray}
By using Eqs. (\ref{C2_100_CFJ}) and (\ref{C_0^100}), 
AMR$^{[100]}(0)$ of Eq. (\ref{AMR^100}) 
is written as
\begin{eqnarray}
\label{100_CFJ}
{\rm AMR}^{[100]}(0) = 2C_2^{[100]} = 
\frac{3}{4} \left( \frac{\lambda}{H} \right)^2 (\alpha -1).
\end{eqnarray}
Equation (\ref{100_CFJ}) is 
the AMR ratio of the CFJ model\cite{Campbell}. 

\item[(2)]${\mbox{\boldmath $I$}}//[110]$ \\
Under the condition of the CFJ model, 
$C_2^{[110]}$ and $C_6^{[110]}$ in Table \ref{table1} become
\begin{eqnarray}
\label{C2_110_CFJ}
&&C_2^{[110]}= \frac{3}{8} \frac{1}{1+r+\alpha} 
\left[ - \left( \frac{\lambda}{H} \right)^2 
\frac{\alpha}{r+\alpha} + 
\left( \frac{\lambda}{H} \right)^2 
\alpha (r + \alpha) \right] \nonumber \\
&& \hspace*{1.1cm}\approx \frac{3}{8} \left( \frac{\lambda}{H} \right)^2 
(\alpha -1),  \\
\label{C6_110_CFJ}
&&C_6^{[110]}=0. 
\end{eqnarray}
By using Eqs. (\ref{C2_110_CFJ}), (\ref{C6_110_CFJ}), and (\ref{C_0^110}), 
AMR$^{[110]}$(0) of Eq. (\ref{AMR^110}) is written as
\begin{eqnarray}
\label{110_CFJ}
{\rm AMR}^{[110]}(0) = 
2(C_2^{[110]} + C_6^{[110]}) = 
\frac{3}{4} \left( \frac{\lambda}{H} \right)^2 (\alpha -1).
\end{eqnarray}
Equation (\ref{110_CFJ}) is 
the AMR ratio of the CFJ model\cite{Campbell}. 

\end{enumerate}


\begin{thebibliography}{9}
\bibitem{Ni_C4}
Substituting 
the experimentally evaluated temperature $T$ dependences of 
$k_1$, $k_2$, $k_3$, and $k_4$ 
[see Fig. 2(a) in Ref. \citen{Dedie}] 
into 
Eqs. (\ref{C2_100_D}), (\ref{C4_100_D}), 
(\ref{C2_110_D}), and (\ref{C4_110_D}), 
we can obtain 
the $T$ dependences of 
$C_2^{[100]}$, $C_4^{[100]}$, 
$C_2^{[110]}$, and $C_4^{[110]}$ for Ni, respectively. 
Namely, 
we have 
($T$, $C_2^{[100]}$, $C_4^{[100]}$, $C_2^{[110]}$, $C_4^{[110]}$)
=(49, 0.0040, 0.0047, 0.0070, $-$0.0047), 
(76, 0.0045, 0.0046, 0.015, $-$0.0046), 
(156, 0.0040, 0.0032, 0.017, $-$0.0032), 
(293, 0.0050, 0.0026, 0.013, $-$0.0026), 
and (360, 0.0025, 0.0010, 0.011, $-$0.0010), 
where the unit of $T$ is K.
\bibitem{[001]}
In the CFJ model, ${\mbox{\boldmath $I$}}$ 
lies in a rotational plane of ${\mbox{\boldmath $M$}}$. 
In this study, 
the cases of ${\mbox{\boldmath $I$}}//[100]$ 
and ${\mbox{\boldmath $I$}}//[110]$ are comparable to the CFJ model, 
whereas the case of ${\mbox{\boldmath $I$}}//[001]$ 
is different from the CFJ model.
\bibitem{HD}
The term with $\lambda^2/(H \Delta)$ in $C_2^{[110]}$ 
is obtained from, for example, 
$\sum_{k (\ne m, m_i)} \sum_{n (\ne m, m_i)}
\frac{ V_{n,m}}{E_m -E_n}
\frac{ V_{k,n}}{E_m -E_k} |k \rangle$ in Eq. (\ref{wf_general}) 
with $E_m -E_n=H$ and $E_m -E_k=\Delta$. 
Here, 
the spin state in $|m \rangle$ 
is the same as that in $|k \rangle$ 
and different from that in $|n \rangle$.
\bibitem{varepsilon12}
For $\rho_{4,\pm}^{[100],(2)}$ of 
Eq. (3) in Ref. \citen{Kokado3_1}, 
we should take into account 
$\rho_{s,\pm \to m,+}=0$ and 
Eqs. (\ref{rho_epsilon1}) and (\ref{rho_epsilon2}).
\bibitem{e-g} 
Under the condition of 
$r_{s,- \to \varepsilon 2,-} = r_{s,- \to \gamma,-}$ 
(i.e., $\rho_{s,- \to \varepsilon 2,-} = \rho_{s,- \to \gamma,-}$), 
we consider 
$\rho_{0,\pm}^{[110],(0)}=
\rho_{s,-} + \frac{3}{4} \rho_{s,- \to \varepsilon 2,-} + 
\frac{1}{4} \rho_{s,- \to \gamma,-}$ of Eq. (\ref{rho_0-^110(0)}), 
where Eq. (\ref{gamma}) is used. 
For this $\rho_{0,\pm}^{[110],(0)}$, 
the above condition may be rewritten as 
$\left| \frac{(3/4)(\rho_{s,- \to \varepsilon 2,-} - \rho_{s,- \to \gamma,-})}{\rho_{s,-}+ \rho_{s,- \to \gamma,-}} \right| \ll 1$. 
When $\rho_{s,- \to \varepsilon 2,-}$ 
is written as 
$\rho_{s,- \to \varepsilon 2,-}= \rho_{s,- \to \gamma,-} + (\rho_{s,- \to \varepsilon 2,-}- \rho_{s,- \to \gamma,-})$, 
we have 
$\rho_{0,\pm}^{[110],(0)}=(\rho_{s,-}+ \rho_{s,- \to \gamma,-})
\left[ 1 + 
\frac{(3/4)(\rho_{s,- \to \varepsilon 2,-} - \rho_{s,- \to \gamma,-})}
{\rho_{s,-}+ \rho_{s,- \to \gamma,-}}\right] 
\approx \rho_{s,-}
+ \rho_{s,- \to \gamma,-}$ 
in the case of 
$\left| \frac{(3/4)(\rho_{s,- \to \varepsilon 2,-} - \rho_{s,- \to \gamma,-})}
{\rho_{s,-}+\rho_{s,- \to \gamma,-}}\right| \ll 1$. 
In this case, 
we obtain 
$\rho_{0,\pm}^{[100],(0)}=\rho_{0,\pm}^{[110],(0)}$ 
of Eq. (\ref{rho_0_relation}). 
Here, $\rho_{0,\pm}^{[100],(0)}$ of 
Eq. (43) in Ref. \citen{Kokado3} 
is given by 
$\rho_{0,\pm}^{[100],(0)} = \rho_{s,\pm} + \rho_{s,\pm \to \gamma,\pm}$, 
where Eq. (\ref{gamma}) is used.
\bibitem{H_Ni}
We report the value of $\lambda/H$ for Ni. 
The spin--orbit coupling constant $\lambda$ 
is evaluated to be $\lambda=8.06 \times 10^{-2}$ eV 
from $\lambda=-(10-n) \lambda_0$, 
with $n$=8 and $\lambda_0 = -4.03 \times 10^{-2}$ eV for Ni$^{2+}$ 
(see Table 1.1 in Ref. \citen{Yosida2}). 
Here, $n$ is the 3d electron number 
and $\lambda_0$ is the spin--orbit coupling constant 
in the spin-orbit interaction 
consisting of 
the total orbital angular momentum and the total spin.\cite{Yosida2} 
Using this $\lambda$ and 
$6.94 \times 10^{-1} \le H \le 7.70 \times 10^{-1}$ eV\cite{Papa}, 
we obtain $1.05 \times 10^{-1} \le \lambda/H \le 1.16 \times 10^{-1}$. 
In this study, we choose 
$\lambda/H = 1.10 \times 10^{-1}$, 
which satisfies the above-mentioned inequality.
\bibitem{rrr}
We chose $r=\rho_{s,-}/\rho_{s,+}=3.00$ 
so as to reproduce the experimental result at 293 K. 
We note that this value is smaller than the theoretical value at 0 K, i.e., 
$r=10.0$ (see $\rho_{s\downarrow}/\rho_{s\uparrow}$ of Ni 
in Table I in Ref. \citen{Kokado1}).
\bibitem{expansion}
We can obtain 
the higher-order terms of 
$C_6^{[110]} \cos 6 \phi'$ and $C_8^{[110]} \cos 8 \phi'$ 
in addition to the D$\ddot{\rm o}$ring expression of Eq. (\ref{Doring_110}) 
by performing the higher-order expansion 
up to the eighth order of the Legendre polynomial 
in the spontaneous magnetostriction theory.\cite{Chikazumi}
\bibitem{fig}
For example, the $\phi$ dependences of 
probabilities and probability amplitudes 
of the specific hybridized states 
are shown in Figs. 4 and 5 in Ref. \citen{Kokado3}.
\end{thebibliography}
\end{document}